\documentclass[11pt]{article}
\pdfoutput=1

\usepackage{jheppubmod}
\usepackage{hyperref}
\usepackage{mathtools}
\usepackage{psfrag}
\usepackage{array}
\usepackage{amssymb}
\usepackage{amsmath}
\usepackage{cancel}
\usepackage{braket}
\allowdisplaybreaks
\usepackage{mathrsfs}
\usepackage{wrapfig}
\usepackage{multicol}
\usepackage{feynmp-auto}

\usepackage{amsthm}
\usepackage{graphicx}
\usepackage{caption}
\usepackage{paralist}
\usepackage{comment}
\usepackage[sorting=none]{biblatex}
\usepackage{dsfont}

\newcommand{\secref}[1]{Section \ref{#1}}
\newcommand{\appref}[1]{Appendix \ref{#1}}
\newcommand{\figref}[1]{Fig. \ref{#1}}

\newcommand{\marrow}[5]{%
\fmfcmd{style_def marrow#1
expr p = drawarrow subpath (1/4, 3/4) of p shifted 6 #2 withpen pencircle scaled 0.4;
label.#3(btex #4 etex, point 0.5 of p shifted 6 #2);
enddef;}
\fmf{marrow#1,tension=0}{#5}}
\newcommand{\Marrow}[6]{%
 \fmfcmd{style_def marrow#1
expr p = drawarrow subpath (1/4, 3/4) of p shifted #6 #2 withpen pencircle scaled 0.4;
label.#3(btex #4 etex, point 0.5 of p shifted #6 #2);
enddef;}
\fmf{marrow#1,tension=0}{#5}}

\title{Charged particle scattering near the horizon}
\author{Fabiano Feleppa$^{\star}$, }
\author{Nava Gaddam$^{\dagger}$, }
\author{Nico Groenenboom$^{\ddag}$} 
\affiliation{$^{\star , \ddag}$Institute for Theoretical Physics and Center for Extreme Matter and Emergent Phenomena,
Utrecht University, 3508 TD Utrecht, The Netherlands.}
\affiliation{$^{\dagger}$International Centre for Theoretical Sciences, Tata Institute for Fundamental Sciences, Bengaluru 560089, Karnataka, India.}
\emailAdd{feleppa.fabiano@gmail.com}
\emailAdd{ n.gaddam@icts.res.in}
\emailAdd{ n.groenenboom@uu.nl}

\date{\today}

\abstract{We study Maxwell theory, in the presence of charged scalar sources, near the black hole horizon in a partial wave basis. We derive the gauge field configuration that solves Maxwell equations in the near-horizon region of a Schwarzschild black hole when sourced by a charge density of a localised charged particle. This is the electromagnetic analog of the gravitational Dray-'t Hooft shockwave near the horizon. We explicitly calculate the S-matrix associated with this shockwave in the first quantised $1\rightarrow 1$ formalism. We develop a theory for scalar QED near the horizon using which we compute the electromagnetic eikonal S-matrix from elastic $2\rightarrow 2$ scattering of charged particles exchanging soft photons in the black hole eikonal limit. The resulting ladder resummation agrees perfectly with the result from the first quantised formalism, whereas the field-theoretic formulation allows for a computation of a wider range of amplitudes. As a demonstration, we explicitly compute sub-leading corrections that arise from four-vertices.}
\begin{document}
\maketitle

\section{Introduction}
Eikonal physics in field theory and gravity arises in the very high energy limit of scattering processes. In field theory, these are $2\rightarrow 2$ elastic $t$-channel scattering processes where external momenta are far greater than virtual momenta that are exchanged \cite{Levy:1969cr}. In perturbative quantum gravity about flat space, these processes involve trans-Planckian scattering where the centre of mass energies of scattering processes satisfy $E \gg M_{Pl}$. Of course, the impact parameter of scattering in this case must necessarily be the largest length scale in the game to remain within the regime of validity of perturbation theory. Eikonal physics in gravitational theories has far reaching theoretical consequences. More recently, its relevance for the calculation of gravitational observables of interest in the inspiral phase of compact binary mergers in gravitational wave astronomy has gained prominence. We refer the reader to a recent review for further details and references \cite{DiVecchia:2023frv}.

Shockwave solutions in flat space \cite{Bonnor:1969, Penrose:1973, Aichelburg:1970dh} are among the first examples that led to eikonal techniques providing for an amplitude-based approach for calculating gravitational observables.\footnote{While the solution of \cite{Bonnor:1969} is not sourced at an instant as one might demand of a shockwave, it has preceded the discovery of shockwaves.} Shockwaves have also been found as non-linear perturbations to classical Einstein equations in black hole backgrounds \cite{Dray:1984ha} and general curved spacetimes \cite{Sfetsos:1994xa}. These shockwaves can be thought to have an eikonal representation in the following two ways. As was argued by 't Hooft, one avatar of these solutions is in terms of the modified (with respect to the background before the insertion of the shockwave) geodesics experienced by observers propagating on these backgrounds. When such an observer is taken to be a probe particle, the classical solution manifests itself as a change in the wavefunction of the said probe particle. This calculation is intrinsically in a first quantised formalism and can be thought of as $1\rightarrow 1$ scattering. This approach was pioneered by 't Hooft both in flat space \cite{tHooft:1987vrq} and in a black hole background \cite{tHooft:1996rdg, Hooft:2016itl, Betzios:2016yaq}.\footnote{See also \cite{Adamo:2021rfq} for a recent application at large distances in curved space.} The second and arguably more powerful avatar is in a field-theoretic setting where the amplitudes arise from elastic $2\rightarrow 2$ scattering of high energy particles exchanging soft virtual modes. This was established in flat spacetime in \cite{Kabat:1992tb}. In the black hole background, the field-theoretic analog is only a recent development. The eikonal manifestation of the Dray-'t Hooft shockwaves in the Schwarzschild black hole in terms of virtual graviton exchange has been developed in \cite{Gaddam:2020rxb, Gaddam:2020mwe}.\footnote{For results in pure AdS and AdS black holes, see \cite{Cornalba:2006xk, Cornalba:2007zb} and \cite{Shenker:2014cwa}, respectively.} In fact, the field-theoretic avatar of eikonal amplitudes in black hole backgrounds has further applications that are difficult to envision in the first quantised $1 \rightarrow 1$ formalism \cite{Betzios:2016yaq, Betzios:2020xuj, Gaddam:2021zka, Gaddam:2022pnb}.

In \cite{tHooft:1996rdg}, 't Hooft argued for an extension of these techniques to including other forces in the Standard Model in the near-horizon region of the black hole. In particular, he argued that the first quantised manifestation of the electromagnetic force near the horizon involves a certain gauge rotation of the gauge field of the charged particle being scattered near the horizon. The primary result of this article is an extension of the techniques developed in \cite{Gaddam:2020rxb, Gaddam:2020mwe} to include charged particle scattering mediated by near-horizon photons. To this end, we organise our presentation as follows: 

\begin{itemize}
	\item We derive the gauge field configuration that solves Maxwell equations in the near-horizon region of a Schwarzschild black hole when sourced by a charge density of a localised charged particle. This is the electromagnetic shockwave analog of the gravitational Dray-'t Hooft shockwave near the horizon. This is done in \secref{sec:EMshockwave}.
	\item In \secref{sec:QMEikonal}, we explicitly calculate the S-matrix associated with this shockwave in the first quantised $1\rightarrow 1$ formalism. Sections \ref{sec:EMshockwave} and \ref{sec:QMEikonal} may be seen as a detailed derivation of the expectations outlined in \cite{tHooft:1996rdg}, and written out in a partial wave basis for direct comparison with the $2\rightarrow 2$ eikonal resummation. 
	\item Finally, we develop a theory for scalar QED near the horizon, following the general formalism of \cite{Gaddam:2020rxb, Gaddam:2020mwe}, in \secref{sec:QEDhorizon}, using which we compute the electromagnetic eikonal S-matrix from elastic $2\rightarrow 2$ scattering of charged particles exchanging soft photons in \secref{sec:QFTEikonal}. The resulting eikonal resummation is identical to the amplitude found in \secref{sec:QMEikonal}, and can be seen as a proof of principle of our proposed effective description for scattering near the horizon of black holes at small impact parameters. Whereas the field-theoretic formulation allows for a computation of a wider range of amplitudes. As a demonstration of the fact, in \secref{app:seagull}, we compute the one-loop diagrams arising from the four-vertex in the theory which are parametrically sub-leading in comparison to the eikonal amplitude that arises from the three-vertex.
\end{itemize}

Our formalism naturally allows for straightforward extensions to non-Abelian gauge fields since we perform these calculations in a basis of partial waves, owed to \cite{Regge:1957td, Gerlach:1980tx}, that has come to much use in the case of scattering of gravitational perturbations in black hole backgrounds. We conclude the paper with further discussion and future directions in \secref{sec:discussion}.

\section{Shockwave of a charged particle in the Schwarzschild background}
In this section, we review 't Hooft's shockwave analysis in the case of a charged particle \cite{tHooft:1996rdg} propagating in the background of a Schwarzschild black hole. The metric for the background, in four dimensions, can be written as follows: 
\begin{equation}\label{eqn:bhmetric}
\mathrm{d}s^{2} ~ = ~ - 2 A\left(u,v\right) \mathrm{d}u \mathrm{d}v + r\left(u,v\right)^{2} \mathrm{d}\Omega^{2}_{(2)} \, ,
\end{equation}
where the functions $A\left(u,v\right)$ and $r\left(u,v\right)$ are defined as
\begin{equation}
A\left(u,v\right) ~ = ~ \dfrac{R}{r} \exp\left(1 - \dfrac{r}{R}\right) \quad \text{and} \quad uv ~ = ~ 2 R^{2} \left(1 - \dfrac{r}{R}\right) \exp\left(\dfrac{r}{R} - 1\right) 
\end{equation}
The line element $\mathrm{d}\Omega^{2}_{(2)}$ defines the round metric on the unit two-sphere and $R = 2 G M$ is the Schwarzschild radius. 

\subsection{Gravitational backreaction and electromagnetic gauge rotation}\label{sec:EMshockwave}
In this section, we review the backreaction of a highly boosted charged shockwave on a probe test particle \cite{Dray:1984ha}. The gravitational backreaction of the shock leaves an imprint on the gravitational field experienced by the probe. The probe then experiences geodesics that are shifted across the null surface traced out by the shockwave.

\subsubsection{Backreaction on the gravitational field}
The stress tensor associated with a localised source carrying momentum $p_{\text{in}}$ at a location $u_{0}$ and a point on the sphere $\Omega_{0}$ can be parametrised as
\begin{align}\label{eqn:gravSource}
T^{\mu \nu} ~ &= ~ - \dfrac{p_{\text{in}}}{A\left(u,v\right) r^{2}} \delta\left(u - u_{0}\right) \delta\left(\Omega - \Omega_{0}\right) \delta^{\mu}_{v} \delta^{\nu}_{v} \nonumber \\
&\sim ~ - \dfrac{p_{\text{in}}}{R^{2}} \delta\left(u - u_{0}\right) \delta\left(\Omega - \Omega_{0}\right) \delta^{\mu}_{v} \delta^{\nu}_{v} \, .
\end{align}
Here, the factor $A\left(u,v\right) r^{2}$ in the first line arises from the square root of the determinant of the longitudinal part of the metric. The angular part is contained in the angular delta function. In the second line, we took a near-horizon limit where $A\left(u,v\right) \sim 1$ when $r \sim R$. An ansatz for the backreacted geometry that solves the Einstein equations with the above source can be taken to be
\begin{align}
\mathrm{d}s^{2} ~ &= ~ - 2 A\left(u,v\right) \mathrm{d} u \Big(\mathrm{d}v - \delta\left(u - u_{0}\right) \lambda_{1}\left(\Omega, \Omega_{0}\right) \mathrm{d} u \Big) + g\left(u,v\right) \mathrm{d}\Omega^{2}_{(2)} \, ,
\end{align}
where, again, $\mathrm{d}\Omega^{2}_{(2)}$ is the line element on the unit round two-sphere. Outside of the location of the source shock, a probe particle experiences the background Schwarzschild solution with a vanishing $\lambda_{1}\left(\Omega, \Omega_{0}\right)$. At the location of the source $\delta\left(u - u_{0}\right)$, however, the Einstein equations with the source \eqref{eqn:gravSource} reduce to \cite{Hooft:2016itl, Betzios:2016yaq}\footnote{In this derivation, terms that are quadratic in $\delta\left(u - u_{0}\right)$ have been neglected. We thank an anonymous referee for bringing ref. \cite{Balasin:1999wg} to our attention that discusses this issue. Our calculation is only valid when the impact parameter between the probe and the shock, measured by the transverse distance on the sphere, is larger than Planck length. This is the regime of validity of this effective description. Beyond this regime, it is of course well-known that a point-particle description in gravity is problematic.}
\begin{align}
\left(\Delta_{\Omega} - 1\right) \lambda_{1}\left(\Omega, \Omega_{0}\right) ~ &= ~ - 8 \pi G \delta \left(\Omega - \Omega_{0}\right) \, ,
\end{align}
where with $\Delta_{\Omega}$ we denoted the Laplacian on the unit two-sphere. Expanding the above equation in partial waves, we find the following solution:
\begin{align}\label{eqn:lambda1}
	\lambda^{\ell m}_{1} ~ = ~ \dfrac{8 \pi G}{\ell^{2} + \ell + 1} \, .
\end{align}

\subsubsection{Backreaction on the electromagnetic field}
In analogy to the gravitational backreaction discussed above, an electromagnetically charged shock leaves an imprint on the electromagnetic field of the probe. The probe then experiences a discontinuity in its electromagnetic field across the null surface traced out by the shockwave. In \cite{tHooft:1996rdg}, 't Hooft argued that this field is pure gauge for a highly boosted observer near the horizon. And therefore, he argued that at the location of the shockwave, its field is affected by the source. To see this explicitly, let us consider a localised source of charge $q_{\text{in}}$, namely
\begin{align}\label{eqn:chargeDensityQM}
J^{\mu} ~ &= ~ \dfrac{q_{\text{in}}}{\sqrt{-g}} \delta\left(u - u_{0}\right) \delta\left(\Omega - \Omega_{0}\right) \delta^{\mu}_{v} \, .
\end{align}
The ansatz for the electromagnetic field of the probe upon the introduction of the above source can be parametrised as 
\begin{align}\label{eqn:shockwaveConfig}
A^{\mu} ~ &= ~ \Theta\left(u - u_{0}\right) \lambda_{2} \left(\Omega , \Omega_{0}\right) \delta^{\mu}_{v} \, .
\end{align}
Therefore, as in the gravitational case, we must solve the Maxwell equations at the location of the horizon $u = 0$:
\begin{align}\label{eqn:eom}
g_{\mu \nu} \Box A^{\nu} -  \nabla_{\mu} \nabla_{\nu} A^{\nu} ~ &= ~ \dfrac{q_{\text{in}}}{\sqrt{-g}} \delta\left(u - u_{0}\right) \delta\left(\Omega - \Omega_{0}\right) \delta^{\mu}_{v} \, .
\end{align}
The left hand side of this equation can be simplified in the Schwarzschild background to find
\begin{align}
\Box A^{\mu} - \partial^{\mu} \left(\nabla \cdot A\right) ~ &= ~ g^{u v} \left(\partial_{u} \tilde{U}_v\right) A^v + 2 \tilde{U}^u  \partial_u A^v + 2 \tilde{V}^u \partial_u A^v \nonumber \\
&\qquad + 2 \tilde{V}^v \tilde{U}_v A^v - 2 \tilde{V}^v \tilde{V}_v A^v + \dfrac{1}{r^2} \Delta_{\Omega} A^v - \partial^v \left[\partial_v \log\left( A r^2\right) A^v \right] \, .
\end{align}
Explicit computation shows that this expression simplifies to reduce \eqref{eqn:eom} to an equation for the undetermined bilocal function $\lambda_{2}\left(\Omega, \Omega_{0}\right)$ in the field configuration \eqref{eqn:shockwaveConfig}:
\begin{equation}
\Delta_{\Omega} \lambda_{2}\left(\Omega, \Omega_{0}\right) ~ = ~ q_{\text{in}} \delta\left(\Omega - \Omega_{0}\right) \, ,
\end{equation}
where we defined $\tilde{V}_{a}\coloneqq \partial_a \log r$ and $\tilde{U}_a \coloneqq \partial_a \log A(r)$. Notice that the Latin index $a$ runs over the coordinates $u$ and $v$. To arrive at this simplification, we integrate the equation against an arbitrary test function to handle the delta function in $u$. In a partial wave basis, we find
\begin{align}\label{eqn:lambda2}
	\lambda^{\ell m}_{2} ~ = ~ - \dfrac{q_{\text{in}}}{\ell \left(\ell + 1\right)} \, .
\end{align}
Therefore, while the electromagnetic field of the probe could be gauge-fixed to vanish in the absence of sources, the backreaction of a source shock results in a gauge rotation of the probe electromagnetic field.

\subsection{An S-matrix for the wavefunction of a probe charged particle}\label{sec:QMEikonal}
The aim of this section is to calculate the S-matrix for the wavefunction of a charged particle in the presence of a gravitationally backreacting charged shockwave. To this end, let us first begin by writing the wavefunction of a charged particle in the said eigenbasis as $\psi \left(p_{\text{in}}, q_{\text{in}}\right) ~ = ~ \langle \psi | p_{\text{in}}, q_{\text{in}} \rangle$. In order to label states as such, we may demand the existence of a charge operator which when acted on its eigenstate yields the charge of the state. Just as a superposition of momentum eigenstates yields a state of definite position, a superposition of charge eigenstates will yield a state with definite electric field. As we argued in the previous subsection, for boosted particles backreacting near the horizon of a black hole, this electric field approaches a pure gauge configuration and may be parameterised by a gauge parameter, say, $\Lambda$. Therefore, we may label states in the momentum-charge basis by $| p , q \rangle$ or by $ | y , \Lambda \rangle$ in the position-gauge field basis. In terms of momentum and charge eigenstates, the S-matrix is formally given by $\left(p_{\text{in}}, q_{\text{in}} ; p_{\text{out}}, q_{\text{out}}\right) ~ \coloneqq ~ \langle p_{\text{in}}, q_{\text{in}} | p_{\text{out}} , q_{\text{out}} \rangle \, .$ This allows us to write
\begin{align}
    \psi \left(p_{\text{in}}, q_{\text{in}}\right) ~ &= ~ \langle \psi | p_{\text{in}}, q_{\text{in}} \rangle \nonumber \\
    &= ~ \int \mathrm{d} q_{\text{out}} \int \frac{\mathrm{d} p_{\text{out}}}{2\pi} \langle \psi | p_{\text{out}}, q_{\text{out}}\rangle \langle p_{\text{out}}, q_{\text{out}} | p_{\text{in}}, q_{\text{in}} \rangle \nonumber \\
    &= ~ \int \mathrm{d} q_{\text{out}} \int \dfrac{\mathrm{d} p_{\text{out}}}{2\pi} \langle \psi | p_{\text{out}}, q_{\text{out}} \rangle S^*\left(p_\text{in}, q_{\text{in}} ; p_\text{out}, q_{\text{out}}\right) \nonumber \\
    &= ~ \int \mathrm{d} \Lambda_{\text{out}} \int \mathrm{d} y \int \mathrm{d} q_{\text{out}} \int \dfrac{\mathrm{d} p_{\text{out}}}{2\pi} \psi \left(y , \Lambda_{\text{out}} \right) \langle y, \Lambda_{\text{out}} | p_{\text{out}}, q_{\text{out}} \rangle \nonumber \\
    &\qquad \qquad \qquad \qquad \qquad \qquad \times ~ S^*\left(p_\text{in}, q_{\text{in}} ; p_\text{out}, q_{\text{out}}\right) \, .
\end{align}
where we used the completeness relations
\begin{align}
   \int \mathrm{d} q_{\text{out}} \int \frac{\mathrm{d} p_\text{out}}{2\pi} | p_\text{out}, q_{\text{out}} \rangle \langle p_\text{out}, q_{\text{out}} | ~ = ~ 1 ~ = ~ \int \mathrm{d} \Lambda_{\text{out}} \int \mathrm{d} y | y , \Lambda_{\text{out}} \rangle \langle y , \Lambda_{\text{out}} | \, ,
\end{align}
and the definition of the scattering matrix. As we argued in the previous section, the gravitational backreaction implies that the position of the outgoing particle is determined by the momentum of the incoming particle. Similarly, the gauge parameter of the outgoing particle is given by the charge of the incoming particle. These relations are\footnote{Here, the ``constants'' $\lambda_{i}$ are only constants along the longitudinal coordinates $u,v$, They indeed depend on the transverse distance on the horizon, between the backreacting shock and the probe outgoing particle.}
\begin{align}\label{eqn:backreaction}
y ~ = ~ \lambda_{1} p_{\text{in}} \quad \text{and} \quad \Lambda_{\text{out}} ~ = ~ \lambda_{2} q_{\text{in}} \, ,
\end{align}
which we insert in the previous expression for the wavefunction to find
\begin{align}\label{eqn:backreactedWavefn}
    \psi \left(p_{\text{in}}, q_{\text{in}}\right) ~ &= ~ \int \lambda_{2} \mathrm{d} q'_{\text{in}} \int \lambda_{1} \mathrm{d} p'_{\text{in}} \int \mathrm{d} q_{\text{out}} \int \dfrac{\mathrm{d} p_{\text{out}}}{2\pi} \psi \left(\lambda_{1} p'_{\text{in}}, \lambda_{2} q'_{\text{in}} \right) \langle y, \Lambda_{\text{out}} | p_{\text{out}}, q_{\text{out}} \rangle \nonumber \\
    &\qquad \qquad \qquad \qquad \qquad \qquad \qquad \times ~ S^*\left(p_\text{in}, q_{\text{in}} ; p_\text{out}, q_{\text{out}}\right) \nonumber \\
    &= ~ \int \mathrm{d} q'_{\text{in}} \int \mathrm{d} p'_{\text{in}} \int \mathrm{d} q_{\text{out}} \int \dfrac{\mathrm{d} p_{\text{out}}}{2\pi} \psi \left(p'_{\text{in}}, q'_{\text{in}} \right) \langle y , \Lambda_{\text{out}} | p_{\text{out}}, q_{\text{out}} \rangle \nonumber \\
    &\qquad \qquad \qquad \qquad \qquad \qquad \qquad \times ~ S^*\left(p_\text{in}, q_{\text{in}} ; p_\text{out}, q_{\text{out}}\right) \, .
\end{align}
The rescaling of integration variables to arrive at the second equality does not change the ranges of integration (which remain from $-\infty$ to $\infty$ for both the integrals.) This relation must hold for any wavefunction as \eqref{eqn:backreaction} contains invertible basis transformations. Therefore, we can finally write
\begin{align}
	\int \mathrm{d} q'_{\text{in}} \int \mathrm{d} p'_{\text{in}} \langle y , \Lambda_{\text{out}} | p_{\text{out}}, q_{\text{out}} \rangle S^*\left(p_\text{in}, q_{\text{in}} ; p_\text{out}, q_{\text{out}}\right) ~ = ~ \delta\left(p'_{\text{in}} - p_{\text{in}}\right) \delta\left(q'_{\text{in}} - q_{\text{in}}\right) \, .
\end{align}
To invert this equation for the S-matrix, we now need an expression for $\langle y , \Lambda_{\text{out}} | p_{\text{out}}, q_{\text{out}} \rangle $. Writing the positions $y$ in a momentum basis gives us a plane wave. Similarly, we know that the electric field and charge density are conjugate and therefore we may write
\begin{align}\label{eqn:basisChange}
    \langle y , \Lambda_{\text{out}} | p_{\text{out}}, q_{\text{out}} \rangle ~ = ~ \exp\left(i \, y \, p_\text{out} + i \Lambda_{\text{out}} q_{\text{out}}\right) ~ = ~ \exp\left(i \, \lambda_{1} \, p_{\text{in}} \, p_\text{out} + i \lambda_{2} \, q_{\text{in}} q_{\text{out}}\right) \, .
\end{align}
Plugging this into the previous expression, we see that it is a Fourier transform equation for the scattering matrix which can easily be inverted to find
\begin{align}
	S^*\left(p_\text{in}, q_{\text{in}} ; p_\text{out}, q_{\text{out}}\right) ~ = ~ \exp\left(- i \, \lambda_{1} \, p_{\text{in}} \, p_\text{out} - i \lambda_{2} \, q_{\text{in}} q_{\text{out}}\right) \, .
\end{align}

\subsection{Generalisation to many particles and the continuum}
We would now like to generalise the previous results to the case of many particles in order to then take a continuum limit to describe a distribution of particles on the horizon. Since quantum mechanics does not allow for particle production, we may safely assume that the number of incoming and outgoing particles is equal and large; we call the number of incoming and outgoing particles as $N_{\text{in}}$ and $N_{\text{out}}$ respectively. We will label the $i$-th incoming particles by its longitudinal position $x_{i}$, angular position on the horizon $\Omega_{i}$, and momentum $p^{i}_{\text{in}}$ such that $i \in N_{\text{in}}$. Similarly, outgoing particles would be labelled by $y_{j}, \Omega_{j}, p^{j}_{\text{out}}$ and $j \in N_{\text{out}}$. Assuming that there is no more than one particle at each angular position on the horizon, in the continuum limit $N_{\text{in}} = N_{\text{out}} \rightarrow \infty$, the positions of particles may be described by distributions $x\left(\Omega\right)$ and $y\left(\Omega\right)$. The basis of states may now be written as
\begin{align}
    | p_{\text{in, tot}} , q_{\text{in, tot}} \rangle ~ = ~ \bigotimes_{i} |p^i_{\text{in}} , q^{i}_{\text{in}} \rangle \quad \text{and} \quad |p_{\text{out, tot}} , q_{\text{out, tot}} \rangle ~ = ~ \bigotimes_{j} | p^j_{\text{out}} , q^{j}_{\text{out, tot}} \rangle \, ,
\end{align}
where we assumed a factorised Hilbert space because all parallel moving particles are independent. The completeness relations are now integrals defined with measure $\mathrm{d} p_\text{out, tot} = \prod_j \mathrm{d} p_\text{out}^j$ and $\mathrm{d} y_\text{tot} = \prod_j \mathrm{d} y^j$. The S-matrix may formally be written as
\begin{align}\label{eqn:smatrixNparticles}
    S_{\text{tot}} ~ \coloneqq ~ S\left(p_{\text{in, tot}}, q_{\text{in, tot}} ; p_{\text{out, tot}}, q_{\text{out, tot}}\right) ~ \coloneqq ~ \langle p_{\text{in, tot}}, q_{\text{in, tot}} | p_{\text{out, tot}} , q_{\text{out, tot}} \rangle \, .
\end{align}
This S-matrix is dictated by the backreaction relations derived before, which are now given in terms of invertible matrices that are in turn functions of the transverse distance between the in and out particles:
\begin{align}
	y^{j} ~ = ~ \lambda^{ij}_{1} \left(\Omega_{i}, \Omega_{j}\right) p^{i}_{\text{in}} \quad \text{and} \quad \Lambda^{j}_{\text{out}} ~ = ~ \lambda^{ij}_{2} \left(\Omega_{i}, \Omega_{j}\right) q^{i}_{\text{in}} \, ,
\end{align}
such that we can write
\begin{align}
	| y_{\text{tot}} , \Lambda_{\text{out, tot}} \rangle ~ = ~ \bigotimes_{j} | \lambda^{ij}_{1} \left(\Omega_{i}, \Omega_{j}\right) p^{i}_{\text{in}} ,  \lambda^{ij}_{2} \left(\Omega_{i}, \Omega_{j}\right) q^{i}_{\text{in}} \rangle \, .
\end{align}
Since the scattering matrix is a basis transformation, it is necessarily bijective between the in and out Hilbert spaces. This implies that the matrices $\lambda_{1}\left(\Omega_{i}, \Omega_{j}\right)$ and $\lambda_{2}\left(\Omega_{i}, \Omega_{j}\right)$ are invertible, which in turn implies that there is no more than one particle entering (leaving) the horizon at any given angle. Moreover, we have the condition that $N_{\text{in}} = N_{\text{out}}$. Consequently, we may now repeat our strategy from the single particle case to the multiparticle case. We begin with the wavefunction
\begin{align}
    \psi \left(p_{\text{in,tot}}, q_{\text{in,tot}}\right) ~ &= ~ \langle \psi | p_{\text{in,tot}}, q_{\text{in,tot}} \rangle \nonumber \\
    &= ~ \int \mathrm{d} q_{\text{out,tot}} \int \frac{\mathrm{d} p_{\text{out,tot}}}{2\pi} \langle \psi | p_{\text{out,tot}}, q_{\text{out,tot}}\rangle \langle p_{\text{out,tot}}, q_{\text{out,tot}} | p_{\text{in,tot}}, q_{\text{in,tot}} \rangle \nonumber \\
    &= ~ \int \mathrm{d} q_{\text{out,tot}} \int \dfrac{\mathrm{d} p_{\text{out,tot}}}{2\pi} \langle \psi | p_{\text{out,tot}}, q_{\text{out,tot}} \rangle S^*_{\text{tot}} \nonumber \\
    &= ~ \int \mathrm{d} \Lambda_{\text{out,tot}} \int \mathrm{d} y_{\text{tot}} \int \mathrm{d} q_{\text{out,tot}} \int \dfrac{\mathrm{d} p_{\text{out,tot}}}{2\pi} \psi \left(y_{\text{tot}} , \Lambda_{\text{out,tot}} \right)  \nonumber \\
    &\qquad \qquad \qquad \qquad \qquad \qquad \times \langle y_{\text{tot}} , \Lambda_{\text{out, tot}} | p_{\text{out, tot}}, q_{\text{out, tot}} \rangle S^*_{\text{tot}} \, ,
\end{align}
where we used the completeness relations
\begin{align}
   \int \mathrm{d} q_{\text{out, tot}} \int \frac{\mathrm{d} p_\text{out, tot}}{2\pi} | p_\text{out, tot}, q_{\text{out, tot}} \rangle \langle p_\text{out, tot}, q_{\text{out, tot}} | ~ &= ~ 1 \, , \\ 
   \int \mathrm{d} \Lambda_{\text{out, tot}} \int \mathrm{d} y_{\text{tot}} | y_{\text{tot}} , \Lambda_{\text{out, tot}} \rangle \langle y_{\text{tot}} , \Lambda_{\text{out, tot}} | ~ &= ~ 1 \, .
\end{align}
We now insert the backreaction relations 
\begin{align}
	y^{j} ~ = ~ \lambda^{ij}_{1} \left(\Omega_{i}, \Omega_{j}\right) p^{i}_{\text{in}} \quad \text{and} \quad \Lambda^{j}_{\text{out}} ~ = ~ \lambda^{ij}_{2} \left(\Omega_{i}, \Omega_{j}\right) q^{i}_{\text{in}} \, ,
\end{align}
resulting in the measures
\begin{align}
	\prod_{j} \mathrm{d} y^{j} ~ = ~ \text{det}\left(\lambda^{ij}_{1} \left(\Omega_{i}, \Omega_{j}\right)\right) \prod_{j} \mathrm{d} p^{i}_{\text{in}} \quad \text{and} \quad \prod_{j} \mathrm{d} \Lambda^{j}_{\text{out}} ~ = ~ \text{det}\left(\lambda^{ij}_{2} \left(\Omega_{i}, \Omega_{j}\right)\right) \prod_{j} \mathrm{d} q^{i}_{\text{in}} \, ,
\end{align}
to write the wavefunction as 
\begin{align}
    \psi \left(p_{\text{in, tot}}, q_{\text{in, tot}}\right) ~ &= ~ \prod_{j} \int  \mathrm{d} q'^{\, i}_{\text{in}} \int  \mathrm{d} p'^{\, i}_{\text{in}} \int \mathrm{d} q_{\text{out, tot}}  \det\left(\lambda^{ij}_{1} \left(\Omega_{i}, \Omega_{j}\right)\right) \det\left(\lambda^{ij}_{2} \left(\Omega_{i}, \Omega_{j}\right)\right) \nonumber \\
    &\qquad \qquad \times \int \dfrac{\mathrm{d} p_{\text{out, tot}}}{2\pi} \psi \left(\lambda^{ij}_{1} p^{i}_{\text{in}} , \lambda^{ij}_{2} q^{i}_{\text{in}} \right)  \nonumber \\
    &\qquad \qquad \qquad \times \langle y_{\text{tot}} , \Lambda_{\text{out, tot}} | p_{\text{out, tot}}, q_{\text{out, tot}} \rangle S^*_{\text{tot}} \, .
\end{align}
For every $j$ in the product, we have a sum over all incoming particles labelled by $i$. In each term of the sum, we rescale the integration variables $p_{\text{in}}$ and $q_{\text{in}}$ to neutralise the corresponding factors of $\lambda_{1}$ and $\lambda_{2}$, just as we did in the single particle case, to arrive at
\begin{align}
    \psi \left(p_{\text{in, tot}}, q_{\text{in, tot}}\right) ~ &= ~ \int \mathrm{d} q'_{\text{in, tot}} \int \mathrm{d} p'_{\text{in, tot}} \int \mathrm{d} q_{\text{out, tot}} \int \dfrac{\mathrm{d} p_{\text{out, tot}}}{2\pi} \psi \left( p'_{\text{in, tot}}, q'_{\text{in, tot}} \right) \nonumber \\
    &\qquad \qquad \qquad \qquad \qquad \qquad \times  \langle y_{\text{tot}} , \Lambda_{\text{out, tot}} | p_{\text{out, tot}}, q_{\text{out, tot}} \rangle S^*_{\text{tot}}  \, .
\end{align}
In analogy to \eqref{eqn:basisChange}, we now write
\begin{align}
\langle y_{\text{tot}} , \Lambda_{\text{out, tot}} | p_{\text{out, tot}}, q_{\text{out, tot}} \rangle ~ &= ~ \prod_{j} \langle y_j , \Lambda^{j}_{\text{out}} | p^{j}_{\text{out}}, q^{j}_{\text{out}} \rangle \nonumber \\
&= ~ \exp\left(i \sum_{j} y_{j} p^{j}_{\text{out}} + i \sum_{j} \Lambda^{j}_{\text{out}} q^{j}_{\text{out}} \right) \, .
\end{align}
Therefore, as we did in the single particle case, we may invert the previous relation for the scattering matrix to finally find
\begin{align}
	S_{\text{tot}} ~ = ~ \exp\left(i \lambda^{ij}_{1} p^{i}_{\text{in}} p^{j}_{\text{out}} + i \lambda^{ij}_{2} q^{i}_{\text{in}} q^{j}_{\text{out}} \right) \, ,
\end{align}
where a sum over all in and out particles is implicit. The continuum limit is now easy to achieve. We first promote the momenta and charges to be distributions as smooth functions of the sphere coordinates and then replace the sum over in and out particles with integrals over the sphere coordinates as
\begin{align}\label{eqn:shockwaveQMSmatrix}
	S_{\text{tot}} ~ &= ~ \exp\left[ i \int \mathrm{d} \Omega \, \mathrm{d} \Omega' \left(\lambda_{1} \left(\Omega, \Omega' \right) p_{\text{in}} \left(\Omega\right) p_{\text{out}} \left(\Omega'\right) + \lambda_{2} \left(\Omega, \Omega' \right) q_{\text{in}} \left(\Omega\right) q_{\text{out}} \left(\Omega' \right) \right) \right] \nonumber \\
	&= ~ \exp\left[ i \left(\dfrac{8 \pi G \, p_{\text{in}} p_{\text{out}}}{\ell^{2} + \ell + 1} + \dfrac{q_{\text{in}} q_{\text{out}}}{\ell \left(\ell + 1\right)} \right) \right] \, ,
\end{align}
where we expanded the expression in partial waves in the second line and substituted for $\lambda_{1}$ and $\lambda_{2}$ using \eqref{eqn:lambda1} and \eqref{eqn:lambda2}. Of course, the momentum and charge distributions are also expanded in spherical harmonics, but their partial wave indices have been suppressed.

\section{Scalar QED near the horizon}\label{sec:QEDhorizon}
In this section, we set up the effective theory of scalar QED near the horizon. Before doing so, it is useful to think about the regime of validity of such an effective theory. As discussed in \cite{Gaddam:2020rxb, Gaddam:2020mwe}, this theory goes beyond semi-classical physics as it takes metric fluctuations into account. This implies that it is a perturbative quantum gravity theory about the black hole background. However, it is important to note that the physics captured by this theory is non-perturbative in nature in comparison to the perturbative quantum gravity theory governed by fluctuations about the global flat vacuum. In the present article, we will restrict our amplitude calculations to only scalar-photon interactions on the black hole background. However, the general theory is meant to include graviton perturbations and should perturbatively be valid to all orders in $\hbar$ including couplings between the photons and gravitons. Nevertheless, it is important to keep in mind that the regime of phase space where the theory is valid is when impact parameters of the scattering particles, say $b$, satisfies $L_{{Planck}} \gg b \geq R$ where $L_{Planck}$ is Planck length and $R$ is the Schwarzschild radius. Since we are interested in near-horizon scattering, the particles inevitably scatter at distances of the order of the Schwarzschild radius or less. Our theory breaks down if impact parameters are sub-Planckian because our effective cannot resolve such short distances. Furthermore, the eikonal diagrams that we will compute are the leading contributions when the centre mass energy of the scattering satisfies $E M_{BH} \gg M^{2}_{Planck}$. Finally, if the energy of scattering, $E$, exceeds the mass of the black hole itself, this regime is analogous to scattering at sub-Planckian distances where the effective low energy description of the theory about a black hole background breaks down. In what follows, we will first derive the effective field theory to set up an appropriate near-horizon limit before explicitly demonstrating the necessity of the restriction on phase space for the validity of our effective theory. Thereafter, we will explicitly compute the diagrams of interest. 

With those general remarks behind us, let us consider a complex scalar field, minimally coupled to the photon in the gravitational background of the Schwarzschild black hole \eqref{eqn:bhmetric}:
\begin{equation}\label{action 2}
S\left[\phi , A_\mu\right] ~ = ~ \int \mathrm{d}^4 x \sqrt{- g} \left[-\left(D_\mu\phi\right)^* \left(D^\mu \phi\right) - m^2 \left| \phi \right|^2 - \dfrac{1}{4}F_{\mu\nu} F^{\mu\nu} \right] \, .
\end{equation}
Here, the covariant derivative $D$ enables gauge and gravitational covariance whereas in what follows, gravitational covariance is enabled by $\nabla$. The action of the former on the complex scalar is defined by
\begin{align}
\left(D_\mu \phi\right)^* \left(D^\mu \phi\right) ~ &= ~ \left(\nabla_\mu \phi - i q A_\mu \phi \right)^* \left(\nabla^\mu \phi - i q A^\mu \phi \right) \nonumber \\
&= ~ \nabla_\mu \phi^* \nabla^\mu \phi + i q A^\mu \left(\phi^* \nabla_\mu \phi - \phi \nabla_\mu \phi^* \right) + q^2 A_\mu^2 \left| \phi \right|^2 \, .
\end{align}
Partial integration now allows us to write the matter action as\footnote{Boundary terms are of course important in general. Given that we will later take a near-horizon limit, it is natural for us to ignore the boundary term at infinity. It is worthwhile to note that this boundary term at infinity is the one that captures the contribution of the flat space eikonal at large distances. However, we are interested in near-horizon physics and a certain eikonalisation at small distances near the horizon. So, one might worry about the contribution of boundary terms that arise from the horizon itself which we also dropped in this equation. At this stage, there is no good reason to do so. As we will later see in the paper, despite having dropped this boundary term on the horizon, the field theory eikonal exactly captures the shockwave amplitude \eqref{eqn:shockwaveQMSmatrix}. It would be interesting to understand the role of potential boundary terms on the horizon, for instance from the perspective of soft theorems near the horizon, but we leave this for future work.}
\begin{align}\label{eqn:matterAction}
S_M ~ &\coloneqq ~ \int \mathrm{d}^4 x \sqrt{-g} \left[ \phi^* \left( \Box - m^2 \right) \phi - q A^\mu j_\mu - q^2 A_\mu^2 \left| \phi \right|^2 \right] \, ,
\end{align}
where with $\Box$ we denoted the d'Alembertian in the Schwarzschild background while the scalar current $j_\mu$ has been defined as
\begin{equation}\label{eqn:current}
j_\mu ~ \coloneqq ~ i \left(\phi^* \nabla_\mu \phi - \phi \nabla_\mu \phi^* \right) \, .  
\end{equation}
The Maxwell action can also be partially integrated to write it in the form $A_\mu \mathcal{O}^{\mu\nu}A_\nu$:
\begin{align}\label{eqn:MaxwellAction}
S_{A_{\mu}} ~ &\coloneqq ~ - \dfrac{1}{4} \int \mathrm{d}^{4} x \left[\left(\nabla_\mu A_\nu - \nabla_\nu A_\mu\right) \left(\nabla^\mu A^\nu -\nabla^\nu A^\mu \right)\right] \nonumber\\
&= ~ \dfrac{1}{2} \int \mathrm{d}^{4} x A_\mu \left[g^{\mu\nu} \Box - \nabla^\mu \nabla^\nu -R^{\mu\nu} \right] A_\nu \, .
\end{align}
Since the Schwarzschild metric is a vacuum solution to Einstein equations, the quadratic operator in \eqref{eqn:MaxwellAction} therefore reduces to
\begin{equation}\label{eqn:quadraticOperPhoton1}
\mathcal{O}^{\mu\nu} ~ \coloneqq ~ g^{\mu\nu} \Box - \nabla^\mu \nabla^\nu \, .
\end{equation}

\subsection{Gauge fixing}

In what follows, we will exploit the background spherical symmetry of the theory to expand the gauge field into partial waves. As Regge and Wheeler argued \cite{Regge:1957td, Ruffini:1972pw}, vector spherical harmonics can be split into even and odd parity modes
\begin{align}
A^{+}_{\mu} \left(u, v, \Omega\right) ~ &= ~ \sum_{\ell m} \begin{pmatrix} A_{a} \left(u, v\right)  \\ A^{+} \left(u, v\right) \partial_{A} \end{pmatrix} Y_{\ell m} \left(\Omega\right) \, , \\ 
A^{-}_{\mu} \left(u, v, \Omega\right) ~ &= ~ \sum_{\ell m} \begin{pmatrix} 0 \\ - A^{-} \left(u, v\right) {\epsilon_{A}}^{B} \partial_{B} \end{pmatrix} Y_{\ell m} \left(\Omega\right) \, ,
\end{align}
where $Y_{\ell m} \left(\Omega\right)$ are the familiar spherical harmonics written in a real basis, lowercase Latin indices represent coordinates along the longitudinal directions and uppercase Latin indices represent coordinates along the transverse sphere. All fields $A_{a}$ and $A^{\pm}$ depend only on longitudinal coordinates and carry partial wave indices which we suppressed to avoid clutter of notation. Moreover, the antisymmetric tensor $ {\epsilon_{A}}^{B}$ whose indices are raised and lowered by the round metric on the sphere or radius $r$ is given by
\begin{align}
 {\epsilon_{A}}^{B} ~ = ~ \begin{pmatrix} 0 & \sin \theta \\ - \csc\theta & 0 \end{pmatrix} \, .
\end{align}
The Maxwell field has a gauge redundancy that needs to be fixed by a choice of gauge. We choose one\footnote{This is achieved by choosing a gauge parameter, say $\Lambda$, such that $\partial_{a} \Lambda = - \partial_{a} A^{+} $, $\partial_{A} \Lambda = - A^{+} \partial_{A}$ and then redefining $A_{a} = A_{a} + \partial_{a} A^{+}$.} where $A^{+} \left(u, v\right) = 0$. This choice may be seen as the adaptation of the ``Regge-Wheeler gauge'' for gravitational scattering to the electromagnetic case. This results in
\begin{align}
A^{+}_{\mu} \left(u, v, \Omega\right) ~ &= ~ \sum_{\ell m} \begin{pmatrix} A_{a} \left(u, v\right) \\ 0 \end{pmatrix} Y_{\ell m} \left(\Omega\right) \, , \label{eqn:evenPhoton} \\ 
A^{-}_{\mu} \left(u, v, \Omega\right) ~ &= ~ \sum_{\ell m} \begin{pmatrix} 0 \\ - A^{-} \left(u, v\right) {\epsilon_{A}}^{B} \partial_{B} \end{pmatrix} Y_{\ell m} \left(\Omega\right) \, . \label{eqn:oddPhoton}
\end{align}
After making the field redefinitions 
\begin{equation}\label{eqn:WeylRescaling}
\mathcal{A}^{\ell m}_{a} \left(u,v\right) ~ \coloneqq ~ \dfrac{\sqrt{A\left(r\right)}  A_{a} \left(u,v\right)}{r\left(u,v\right)} \quad \text{and} \quad \mathcal{A}^{\ell m}\left(r\right) ~ \coloneqq ~ \dfrac{A^{-}\left(u,v\right)}{r\left(u,v\right)} \, ,
\end{equation}
and plugging the spherical harmonic decomposition \eqref{eqn:evenPhoton} and \eqref{eqn:oddPhoton} into the Maxwell action, Eq. \eqref{eqn:MaxwellAction}, we find 
\begin{align}
S_{A^{+}_{\mu}} ~ = ~ \dfrac{1}{2} \sum_{\ell m} \int \mathrm{d}^{2} x \, \mathcal{A}^{a}_{\ell m} \Delta^{-1}_{a b}  \mathcal{A}^{b}_{\ell m} \quad \text{and} \quad S_{A^{-}_{\mu}} ~ &= ~ \dfrac{1}{2} \sum_{\ell m} \int \mathrm{d}^{2} x \,  \mathcal{A}_{\ell m} \Delta^{-1} \mathcal{A}_{\ell m} \, .
\end{align}
Above, the quadratic operators are given by
\begin{subequations}\label{eqn:generalQuadOpers}
\begin{align}
\Delta_{ab}^{-1} ~ &= ~ \eta_{ab} \left\{ \eta^{c d} \partial_{c} \partial_{d} -\dfrac{A\left(r\right)^2}{16 r^2 R^2} \left[ \left( 1+ \dfrac{r}{R} \right)^2 + 2 \left( 1 + \dfrac{r}{R}\right) - 8 \left( 2 + \dfrac{r}{R} \right) + 4 \right] x_a x^a - \dfrac{A\left(r\right)}{rR} \right. \nonumber \\
& \qquad \qquad \left. -\hspace{1mm} \dfrac{\left(\lambda - 1\right) A\left(r\right)}{r^2} + \dfrac{A\left(r\right)}{4 r R} \left( 1 + \dfrac{r}{R} \right) - \dfrac{A\left(r\right) R}{2 r^3}\right\} - \dfrac{A\left(r\right)}{4 r R} \left( 3 + \dfrac{r}{R} \right) \left(x_b \partial_a - x_a \partial_b \right) \nonumber \\
& \qquad \qquad + \dfrac{A\left(r\right)^2}{16 r^2 R^2} \left[ \left( 1 + \dfrac{r}{R} \right)^2 - 2 \left( 2 + 2 \dfrac{r}{R} + \dfrac{r^2}{R^2} \right) \right] x_a x_b  - \partial_{a} \partial_{b}\label{eqn:generalEvenOper} \, , \\
\Delta^{-1} ~ &= ~ \left(\lambda - 1\right) \eta^{a b} \partial_{a} \partial_{b} - \dfrac{A\left(r\right) \left(\lambda - 1\right) \left( \lambda - 2\right)}{r^{2}} - \dfrac{2 \left(\lambda - 1\right) A\left(r\right) R}{r^{3}} \nonumber \\
	&\qquad \qquad - \dfrac{A\left(r\right) \left(\lambda - 1\right) }{ r R} x^{a} \partial_{a} \, , \label{eqn:generalOddOper}
\end{align}
\end{subequations}
where we defined $\lambda \coloneqq \ell^{2} + \ell + 1$ and $\eta_{ab}$ is the flat metric in two-dimensions with off-diagonal elements given by $-1$. It is evident that we have traded a single four-dimensional theory in the Schwarzschild background for an infinite tower of decoupled two-dimensional theories, one for each partial wave, with curvature effects encapsulated in potentials. We present the details of this calculation in \appref{app:MaxwellActionPW}.

\subsection{Near horizon limits and the photon propagator}
While the four-dimensional theory in curved space can be simplified into decoupled two-dimensional theories in flat space with extra potentials as we demonstrated in the previous section, we have not lost any generality. Therefore, it is still an analytically intractable task to invert the quadratic operators \eqref{eqn:generalQuadOpers}. As was shown in the gravitational case in \cite{Gaddam:2020rxb, Gaddam:2020mwe}, the way forward is a near-horizon approximation where the operators simplify considerably. 

\subsubsection{Shockwave approximation}\label{sec:shockwaveApprox}
Since the eikonal approximation near the black hole horizon derived its motivation from consideration of shockwave geometries, it is natural to impose a constraint on the gauge field fluctuations to obey the shockwave configuration \eqref{eqn:shockwaveConfig} in the near-horizon region. We would like to impose these restrictions in a covariant manner on the longitudinal directions for each partial wave. Considering a near-horizon approximation to linear order (in $u, v$) implies that $x^{a} x_{a} \sim u v \sim 0$. Additionally, the shockwave approximation can be captured by the condition that $x^{a} A_{a} \sim u A_{u} + v A_{v} \sim 0$. This is understood as follows. Consider the past horizon located at $v=0$. Thus, one of the terms\footnote{Notice that in comparison to \eqref{eqn:shockwaveConfig}, a lowering of the gauge field index is achieved by the metric as $A_{u} = g_{u v} A^{v}$. Near the horizon, $g_{u v} \sim -1$. }, $v A_{v}$, is naturally vanishing on the horizon. The second term drops if we choose the shockwave configuration as in \eqref{eqn:shockwaveConfig} where the $u$-component of the gauge field vanishes. An analogous approximation clearly holds on the past horizon. 

In order to employ this approximation on \eqref{eqn:generalQuadOpers}, we first note that to linear order (in $u, v$), $r\left(u,v\right) = R + \mathcal{O}\left(uv\right)$ and therefore, $A\left(u,v\right) \sim 1$. With these considerations, the quadratic operators \eqref{eqn:generalQuadOpers} simplify to
\begin{subequations}
\begin{align}
\Delta_{ab}^{-1} ~ &= ~ \eta_{ab} \left( \eta^{c d} \partial_{c} \partial_{d} - \dfrac{\lambda}{R^2} \right) - \dfrac{x_b \partial_a}{R^{2}} - \partial_{a} \partial_{b} \, , \\
\Delta^{-1} ~ &= ~ \left(\lambda - 1\right) \eta^{a b} \partial_{a} \partial_{b} - \dfrac{\lambda \left( \lambda - 1\right)}{R^{2}} - \dfrac{\left(\lambda - 1\right) }{ R^{2} } x^{a} \partial_{a} \, .
\end{align}
\end{subequations}
In the action, the term containing $x_{b} \partial_{a}$ can further be simplified as
\begin{align}
	\int \mathrm{d}^2 x \mathcal{A}^a x_b \partial_a \mathcal{A}^b ~ &= ~ \int \mathrm{d}^{2} x \left( \partial_a \left(\mathcal{A}^a x_b \mathcal{A}^b \right) - \mathcal{A}^a \eta_{bc}\partial_a x^c \mathcal{A}^b\right) ~ = ~ - \int \mathrm{d}^{2} x\mathcal{A}^a \eta_{ab}\mathcal{A}^b \, .
\end{align}
Similarly, the term containing $x^{a} \partial_{a}$ can also be simplified to
\begin{align}
	- \dfrac{\left(\lambda - 1\right) }{ R^{2} } \int \mathrm{d}^2 x \mathcal{A} x^b \partial_b \mathcal{A} ~ &= ~ - \dfrac{\left(\lambda - 1\right) }{2 R^{2} } \int \mathrm{d}^2 x \, x^b \partial_b \mathcal{A}^2 \nonumber \\
	&= ~ \dfrac{\left(\lambda - 1\right) }{R^{2} } \int \mathrm{d}^2 x \, \mathcal{A}^2 + \dfrac{\left(\lambda - 1\right) }{R^{2} } \int \mathrm{d}^2 x \, \partial_{b} \left( x^{b} \mathcal{A}^{2} \right) \, .
\end{align}
The issue of the boundary term may potentially be subtle. On first glance, it appears to be safe to assume that the field falls off at the boundaries. However, notions of ``far past'' and ``far future'' on the horizon are not well-defined in an evaporating black hole formed by collapse within the effective field theory regime being considered in this paper. Nevertheless, we will blithely ignore this boundary term in this work and leave a careful analysis of the relevance of it in the effective theory for the future. Therefore, the quadratic operators in this approximation scheme can be written in their final form in the following way
\begin{subequations}\label{eqn:shockwaveQuadOpers}
\begin{align}
\Delta_{ab}^{-1} ~ &= ~ \eta_{ab} \left( \eta^{c d} \partial_{c} \partial_{d} - \dfrac{\left(\lambda - 1\right)}{R^2} \right) - \partial_{a} \partial_{b} \, , \\
\Delta^{-1} ~ &= ~ \left(\lambda - 1\right) \left[\eta^{a b} \partial_{a} \partial_{b} - \dfrac{ \left( \lambda - 1\right)}{R^{2}} \right] \, .
\end{align}
\end{subequations}
It is noteworthy that when $\ell = 0$, we have that $\lambda = 1$ and thus the odd action vanishes. This is consistent with the fact that there are no odd degrees of freedom in the monopole sector.

\paragraph{Propagator for the photon:} These quadratic operators above, in Eq. \eqref{eqn:shockwaveQuadOpers}, may be written in Fourier space as follows:
\begin{align}
\Delta_{ab}^{-1}\left(p\right) ~ &= ~ - \eta_{ab} \left( p^{2} + \dfrac{\left(\lambda - 1\right)}{R^2} \right) + p_{a} p_{b} \, , \\
\Delta^{-1}\left(p\right) ~ &= ~ - \left(\lambda - 1\right) \left[ p^{2} + \dfrac{ \left( \lambda - 1\right)}{R^{2}} \right] \, .
\end{align}
In order to find their inverses, we demand that
\begin{align}
	\Delta_{a b}^{-1}\left(p\right) \Delta^{b c}\left(p'\right) ~ = ~ \delta_{a}^{c} \, \delta^{(2)} \left(p - p'\right) \quad \text{and} \quad \Delta^{-1}\left(p\right) \Delta\left(p'\right) ~ = ~  \delta^{(2)} \left(p - p'\right) \, .
\end{align}
Lorentz invariance along the longitudinal directions near the horizon implies that the most general ansatz for the propagator for the even mode can be written as
\begin{align}
	\Delta^{b c}\left(k\right) ~ = ~ f_{1}\left(k^{2}\right) \left(\eta^{ b c } - f_{2}\left(k^{2}\right) k^{b} k^{c}\right) \, .
\end{align}
Explicitly computing $\Delta_{a b}^{-1}\left(k\right) \Delta^{b c}\left(k'\right)$ and solving for the unknown functions $f_{i}$, we find that the propagator for the even mode of the photon is
\begin{equation}\label{eqn:evenPhotonProp_shockwaveApprox}
	\Delta^{a b}\left(k\right) ~ = ~ \dfrac{-1}{k^{2} + \frac{\lambda - 1}{R^{2}} - i \epsilon} \left(\eta^{ab} + \dfrac{R^{2} k^{a} k^{b}}{\lambda - 1}\right) \, . 
\end{equation}
In similar vein, the propagator for the odd mode of the photon can be worked out to be
\begin{align}\label{eqn:oddPhotonProp_shockwaveApprox}
	\Delta\left(k\right) ~ = ~ \dfrac{ - 1}{\left(\lambda - 1\right) \left[ p^{2} + \frac{ \left( \lambda - 1\right)}{R^{2}} \right]} \, .
\end{align}
Just as in the case of the graviton, the photon acquires an effective mass near the horizon owing to curvature effects, while the photon in four dimensions remains massless as it must.

\subsubsection{A leading order near-horizon approximation}\label{sec:leadingOrderApprox}
As it turns out, there is a different approximation that simplifies the quadratic operators considerably. This was also noted in the case of the graviton \cite{Gaddam:2020mwe}. In this approximation, unlike in the shockwave approximation, the configurations that the photon may acquire are not constrained. Instead, we simply work to leading order in the near-horizon approximation assuming that the gauge field does not blow up on the horizon. This implies that all terms proportional to $x^{a}$ can be dropped, leading us to the following operators in this scheme:
\begin{subequations}\label{eqn:leadingOrderApproxQuadOpers}
\begin{align}
\Delta_{ab}^{-1} ~ &= ~ \eta_{ab} \left( \eta^{c d} \partial_{c} \partial_{d} - \dfrac{\lambda}{R^2} \right) - \partial_{a} \partial_{b} \, , \\
\Delta^{-1} ~ &= ~ \left(\lambda - 1\right) \left[\eta^{a b} \partial_{a} \partial_{b} - \dfrac{\lambda}{R^{2}}\right] \, .
\end{align}
\end{subequations}
Following the calculation in the shockwave approximation, the corresponding propagators in this alternative leading order near-horizon approximation can easily be found to be
\begin{align}
	\Delta^{a b}\left(k\right) ~ &= ~ \dfrac{-1}{k^{2} + \frac{\lambda}{R^{2}} - i \epsilon} \left(\eta^{ab} + \dfrac{R^{2} k^{a} k^{b}}{\lambda}\right) \, , \\
	\Delta\left(k\right) ~ &= ~ \dfrac{ - 1}{\left(\lambda - 1\right) \left( p^{2} + \frac{\lambda}{R^{2}} \right)} \, . 
\end{align}

\subsection{Interaction vertices}
In this section, we proceed with writing the interaction vertices in a partial wave basis, starting from the three-vertex in the following section and subsequently focussing on the four-vertex.

\subsubsection{Three-point interaction}\label{sec:threeVertex}

The three-vertex in \eqref{eqn:matterAction} is given by
\begin{align}
	S^{(3)} ~ &\coloneqq ~ - i q \int \mathrm{d}^{4} x \sqrt{-g} \, A^{\mu} \left(\phi^{\star} \nabla_{\mu} \phi - \phi \nabla_{\mu} \phi^{\star}\right) \nonumber \\
	&= ~ - i q \int \mathrm{d}^4 x \sqrt{-g} \, \left[ g^{ab} A_a \left(\phi^\star \partial_b \phi -\phi \partial_b \phi^\star \right)+g^{AB}A_A \left(\phi^\star \partial_B \phi -\phi \partial_B \phi^\star \right)\right] \, .
\end{align}
Since the dominant contribution to the high-energy amplitudes in the eikonal sector arise from the longitudinal momenta, we will henceforth ignore the transverse effects. This amounts to dropping the second term in the square brackets above. We then expand all fields in partial wave basis to find
\begin{align}
	S^{(3)} ~ &= ~ - i q \sum_{\substack{\ell m \\ \ell_{1} m_{1} \\ \ell_{2} m_{2}}} \int \mathrm{d}^{2}x A\left(r\right) r^{2} g^{ab} A^{\ell m}_{a} \left(\phi^\star_{\ell_{1} m_{1}} \partial_b \phi_{\ell_{2} m_{2}} -\phi_{\ell_{2} m_{2}} \partial_b \phi^\star_{\ell_{1} m_{1}} \right) C^{(3)}\left[\ell_{i} , m_{i}\right] \, .
\end{align}
To arrive at this expression, we defined the following integral of three spherical harmonics at different $\ell, m$'s on the two-sphere:
\begin{equation}
	C^{(3)}\left[\ell_{i} , m_{i}\right]  ~ \coloneqq ~ \int \mathrm{d}\Omega_{(2)} Y_{\ell m}\left(\Omega\right) Y_{\ell_{1} m_{1}}\left(\Omega\right) Y_{\ell_{2} m_{2}}\left(\Omega\right) \, .
\end{equation}
In general, interaction terms break the spherical symmetry of the background as can be seen from the presence of the Clebsch-Gordon coefficients in the three-vertex above. While it is certainly possible to perform several calculations with this general vertex, it turns out to be very cumbersome for the resummation of eikonal diagrams. Therefore, it is convenient to choose one scalar leg in each vertex of the diagrams to always be in a fixed partial wave, say $\ell = 0$. Such a choice may be thought of as being reasonable given that we do not imagine the spherical symmetry of the large black hole background to be badly destroyed by perturbative scattering processes. This approximation then leads us to a simplification of the above three-vertex where one of the spherical harmonics merely gives an overall factor of $Y_{00}$ as follows:
\begin{align}
	S^{(3)} ~ &\approx ~ - i \dfrac{q}{\sqrt{4 \pi}} \sum_{\ell m} \int \mathrm{d}^{2}x A\left(r\right) r^{2} g^{ab} A^{\ell m}_{a} \left(\phi^\star_{\ell m} \partial_b \phi_{0} -\phi_{0} \partial_b \phi^\star_{\ell m} \right) \, ,
\end{align}
where we denoted the scalar mode in the s-wave by $\phi_{0}$. In order to use the same photon mode that appeared in the propagators of the previous section, we perform the field redefinition in Eq. \eqref{eqn:WeylRescaling} in addition to rescaling the scalar field as $\phi \rightarrow \frac{\varphi}{r}$ to find
\begin{align}
	S^{(3)} ~ &\approx ~ - i \dfrac{q}{\sqrt{4 \pi} \, R} \sum_{\ell m} \int \mathrm{d}^{2}x \, \eta^{ab} \mathcal{A}^{\ell m}_{a} \left(\varphi^\star_{\ell m} \partial_b \varphi_{0} - \varphi_{0} \partial_b \varphi^\star_{\ell m} \right) \, .
\end{align}
This result is approximate in two ways. One is that we have ignored the mixing of partial waves as described above. On the other hand, we took a near-horizon limit where the field redefinitions of the scalar result in sub-leading terms in $1/R$ which we have ignored.

\subsubsection{Four-point interaction}\label{sec:fourVertex}
Next, we move to the four vertex in \eqref{eqn:matterAction}:
\begin{align}
	S^{(4)} ~ &\coloneqq ~ - q^{2} \int \mathrm{d}^{4} x \sqrt{-g} \, A^{\mu} A_{\mu} \left|\phi^{2}\right| \nonumber \\
	&= ~ - q^2 \int \mathrm{d}\Omega \int \mathrm{d}^2 x A\left(r\right) r^2 \left[g^{ab} A_a A_b + g^{AB} A_A A_B \right] \left|\phi\right|^2 \, .
\end{align}
We now expand all fields in partial waves as before. However, the integral over the two-sphere now involves four spherical harmonics in the even sector. Whereas in the odd sector, two of the four spherical harmonics come with derivatives on them as can be seen from the definition of the odd component of the photon in \eqref{eqn:oddPhoton}. Following our previous choice to ignore partial wave mixing, we now take both scalar modes in the vertex to be in the s-wave.\footnote{In the odd sector, there is no other available choice since the odd mode of the photon vanishes identically in the s-wave. The even sector, however, allows for more choice but we make the simplest one. Other choices, or even the most general integral with all four spherical harmonics, may just as well be written down in terms of products of Clebsch-Gordon coefficients.} Finally, redefining the fields as in the three-vertex case, we find 
 \begin{align}
 	S^{(4)} ~ &\approx ~ - \dfrac{q^{2}}{4 \pi R^{2}} \int \mathrm{d}^{2} x \left(\mathcal{A}^{2}_{a, \ell m} + \left(\lambda - 1\right) \mathcal{A}^{2}_{\ell m}\right) \left|\varphi\right|^{2} \, .
 \end{align}
 To arrive at this expression, in the odd sector, we made use of the following familiar integral
 \begin{align}
 	\int \mathrm{d}\Omega \, \epsilon^{A B} D_{B} Y_{\ell m} {\epsilon_{A}}^{C} D_{C} Y_{\ell' m'} ~ = ~ \left(\lambda - 1\right) \delta_{\ell \ell'} \delta_{m m'} \, .
 \end{align}

\section{Eikonal S-matrix = shockwave S-matrix}\label{sec:QFTEikonal}
Having built up all the tools necessary for computing scattering amplitudes in the theory near the black hole horizon, we first summarise the necessary Feynman rules before moving on to the computations of the amplitudes.

\subsection{Feynman rules near the horizon}\label{sec:FeynRules}
In this section, we collect all the Feynman rules we have derived in the previous sections (propagator of the even mode, propagator of the odd mode, scalar propagator and finally the two vertices).
\begin{itemize}
	\item The propagators of the even mode of the photon in the shockwave and leading order near-horizon approximations, respectively, are
	\begin{subequations}
		\begin{align}
			\mathcal{P}^{a b}\left(k\right) ~ &= ~ \dfrac{-i}{k^{2} + \frac{\lambda - 1}{R^{2}} - i \epsilon} \left(\eta^{ab} + \dfrac{R^{2} k^{a} k^{b}}{\lambda - 1}\right) \label{eqn:evenPhotonProp_shockwaveApprox} \, , \\
			\mathcal{P}^{a b}\left(k\right) ~ &= ~ \dfrac{-i}{k^{2} + \frac{\lambda}{R^{2}} - i \epsilon} \left(\eta^{ab} + \dfrac{R^{2} k^{a} k^{b}}{\lambda}\right) \label{eqn:evenPhotonProp_leadingOrderApprox} \, .
		\end{align}
	\end{subequations}
	\item The propagators of the odd mode of the photon, on the other hand, again in the shockwave and leading order near-horizon approximations, respectively, are given by
	\begin{subequations}
		\begin{align}
			\mathcal{P}\left(k\right) ~ &= ~ \dfrac{ - i}{\left(\lambda - 1\right) \left( p^{2} + \frac{\lambda - 1}{R^{2}} - i \epsilon \right)}  \label{eqn:oddPhotonProp_shockwaveApprox} \, , \\
			\mathcal{P}\left(k\right) ~ &= ~ \dfrac{ - i}{\left(\lambda - 1\right) \left( p^{2} + \frac{\lambda}{R^{2}} - i \epsilon \right)} \label{eqn:oddPhotonProp_leadingOrderApprox} \, .
		\end{align}
	\end{subequations}
\end{itemize}
These were derived in \secref{sec:shockwaveApprox} and \secref{sec:leadingOrderApprox}. 
\begin{itemize}
	\item The scalar propagator is straightforward to compute and was done in \cite{Gaddam:2020mwe}. We have:
		\begin{equation}
			\mathcal{P}_{\phi}\left(k\right) ~ = ~ \dfrac{- i}{k^{2} + \frac{\lambda}{R^{2}} + m^{2} - i \epsilon} \, . \label{eqn:scalarProp}
		\end{equation}
	\item Next, we have two three-vertices arising from the results of \secref{sec:threeVertex}. These are drawn in \figref{fig:threeVertex} below.
 \vspace{0.3cm}
		\begin{figure}[h!]
			\centering
				\begin{multicols}{2}
					\begin{fmffile}{feyn3vertex1}
						\begin{align}
							\hspace{-1.5cm}
							\begin{fmfgraph*}(80,80)
							\fmfleft{i8,i7}
							\fmfright{o2}
							\fmfv{label=$\hat{\varphi}_0 (p_1)$,label.angle=90,label.dist=6}{i7}
							\fmfv{label=$\hat{\varphi}^* (-p_2)$,label.angle=90,label.dist=-17}{i8}
							\fmflabel{$\hat{\mathcal{A}}^b (k) \hspace{0.3cm} = \hspace{0.3cm} \dfrac{i q}{R \sqrt{4\pi}}\left(p_b^1 +p_b^2 								\right) \hspace{0.2cm} =  \hspace{2cm} $}{o2}
							\fmf{dashes_arrow,tension=1.5}{v1,i7}
							\fmf{fermion,tension=1.5}{i8,v1}
							\fmf{wiggly}{v1,o2}  
							\fmffreeze
							\marrow{a}{right}{rt}{$p_1$}{v1,i7}
							\marrow{b}{right}{rt}{$p_2$}{i8,v1}
							\marrow{c}{up}{top}{$k$}{v1,o2}
							\end{fmfgraph*}\nonumber
						\end{align}
					\end{fmffile}
					\begin{fmffile}{feyn3vertex2}
						\begin{align}
							\begin{fmfgraph*}(80,80)
							\fmfleft{i8,i7}
							\fmfright{o2}
							\fmfv{label=$\hat{\varphi}_0^* (-p_1)$,label.angle=90,label.dist=6}{i7}
							\fmfv{label=$\hat{\varphi}(p_2)$,label.angle=90,label.dist=-17}{i8}
							\fmflabel{$\hat{\mathcal{A}}^b (k) $}{o2}
							\fmf{dashes_arrow,tension=1.5}{i7,v1}
							\fmf{fermion,tension=1.5}{v1,i8}
							\fmf{wiggly}{v1,o2}  
							\fmffreeze
							\marrow{a}{right}{rt}{$p_1$}{i7,v1}
							\marrow{b}{right}{rt}{$p_2$}{v1,i8}
							\marrow{c}{up}{top}{$k$}{v1,o2}
							\end{fmfgraph*}\nonumber
						\end{align}
					\end{fmffile}
				\end{multicols}
                \vspace{0.2cm}
				\caption{Here, the dashed lines refer to the scalar mode in the s-wave whereas the solid black line corresponds to the scalar mode in an arbitrary partial wave. The hats indicate that the modes are in Fourier space. The arrows superimposed on the scalar legs indicate the flow of charge while the external arrows indicate flow of momentum.}
				\label{fig:threeVertex}
		\end{figure}
		\item Finally, we have two four vertices, one each from the even and odd photon as can be seen from \secref{sec:fourVertex}. These are drawn in \figref{fig:fourVertex} below.
		\begin{figure}[ht]
			\centering
				\begin{multicols}{2}
					\begin{fmffile}{feyn4vertex1}
						\begin{align}
						\hspace{-2cm}
							\begin{fmfgraph*}(70,70)
							\fmfleft{i1}
							\fmfright{o1}
							\fmfv{label=$\hat{\varphi}_0^* (p_1)$}{i1}
							\fmfv{label=$\hspace{-0.1cm}\hat{\varphi}_0 (p_2)$}{o1}
							\fmfv{label=$\hat{A}^a (k)$}{t1}
							\fmfv{label=$\hat{A}^b (k^\prime)$}{t2}
							\fmflabel{$\vspace{1.2cm}\hspace{2.5cm} = - \dfrac{2 i q^2}{4 \pi R^{2}}\eta_{ab} ,$}{v1}
							\fmf{dashes_arrow}{i1,v1,o1}
							\fmffreeze   
							\fmftop{t1,t2}
							\fmf{photon}{t1,v1}
							\fmf{photon}{t2,v1}
							\marrow{a}{down}{bot}{$p_1$}{i1,v1}
							\marrow{b}{down}{bot}{$p_2$}{o1,v1}
							\marrow{c}{up}{top}{$k$}{v1,t1}
							\marrow{d}{up}{top}{$k^\prime$}{v1,t2}
							\end{fmfgraph*}\nonumber
						\end{align}
					\end{fmffile}

					\begin{fmffile}{feyn4vertex2}
						\begin{align}
						\hspace{-1.5cm}
							\begin{fmfgraph*}(70,70)
							\fmfleft{i1}
							\fmfright{o1}
							\fmfv{label=$\hat{\varphi}_0^* (p_1)$}{i1}
							\fmfv{label=$\hspace{-0.1cm}\hat{\varphi}_0 (p_2)$}{o1}
							\fmfv{label=$\hat{A}(k)$}{t1}
							\fmfv{label=$\hat{A}(k^\prime)$}{t2}
							\fmfv{label=$\vspace{1.2cm}\hspace{2.5cm} = - \dfrac{2 i q^2}{4 \pi R^{2}} .$}{v1}
							\fmf{dashes_arrow}{i1,v1,o1}
							\fmffreeze   
							\fmftop{t1,t2}
							\fmf{wiggly,foreground=blue}{t1,v1}
							\fmf{wiggly,foreground=blue}{t2,v1}
							\marrow{a}{down}{bot}{$p_1$}{i1,v1}
							\marrow{b}{down}{bot}{$p_2$}{o1,v1}
							\marrow{c}{up}{top}{$k$}{v1,t1}
							\marrow{d}{up}{top}{$k^\prime$}{v1,t2}
							\end{fmfgraph*}\nonumber
						\end{align}
					\end{fmffile}
				\end{multicols}
				\vspace{-1cm}
				\caption{In these vertices, the solid black wiggly lines represent the even mode of the photon as in the three-vertex case, whereas the blue wiggly line refers to the odd mode of the photon. The scalar modes remain as before.}
				\label{fig:fourVertex}
		\end{figure}
\end{itemize}

\subsection{Tree level elastic $2-2$ diagrams}
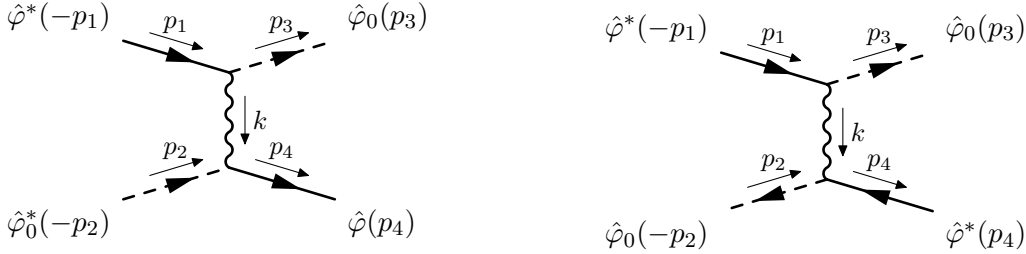
\begin{figure}[h!]
	\centering
	\begin{multicols}{2}
		\begin{fmffile}{feynelectronelectron}
			\begin{align}
				\begin{fmfgraph*}(80,60)
				\fmfstraight
				\fmfleft{i1,i2}
				\fmfright{o1,o2}
				\fmflabel{$\hat{\varphi}_0^* (-p_2)$}{i1}
				\fmfv{label=$\hat{\varphi}(p_4)$}{o1}
				\fmflabel{$\hat{\varphi}_0 (p_3)$}{o2}
				\fmflabel{$\hat{\varphi}^* (-p_1)$}{i2}
				\fmf{dashes_arrow,tension=1.5}{i1,v1}
				\fmf{fermion,tension=1.5}{v1,o1}
				\fmf{photon}{v1,v2}
				\fmf{fermion,tension=1.5}{i2,v2}
				\fmf{dashes_arrow,tension=1.5}{v2,o2}
				\marrow{a}{up}{top}{$p_1$}{i2,v2}
				\marrow{c}{up}{top}{$p_2$}{i1,v1}
				\marrow{b}{up}{top}{$p_3$}{v2,o2}
				\marrow{d}{up}{top}{$p_4$}{v1,o1}
				\marrow{e}{right}{rt}{$k$}{v2,v1}
				\end{fmfgraph*}\nonumber
			\end{align}
		\end{fmffile}

		\begin{fmffile}{feynelectronpositron}
			\begin{align}
				\begin{fmfgraph*}(80,60)
				\fmfstraight
				\fmfleft{i1,i2}
				\fmfright{o1,o2}
				\fmflabel{$\hat{\varphi}_0 (-p_2)$}{i1}
				\fmfv{label=$\hat{\varphi}^* (p_4)$}{o1}
				\fmflabel{$\hat{\varphi}_0 (p_3)$}{o2}
				\fmflabel{$\hat{\varphi}^* (-p_1)$}{i2}
				\fmf{dashes_arrow,tension=1.5}{v1,i1}
				\fmf{fermion,tension=1.5}{o1,v1}
				\fmf{photon}{v1,v2}
				\fmf{fermion,tension=1.5}{i2,v2}
				\fmf{dashes_arrow,tension=1.5}{v2,o2}
				\marrow{a}{up}{top}{$p_1$}{i2,v2}
				\marrow{c}{up}{top}{$p_2$}{i1,v1}
				\marrow{b}{up}{top}{$p_3$}{v2,o2}
				\marrow{d}{up}{top}{$p_4$}{v1,o1}
				\marrow{e}{right}{rt}{$k$}{v2,v1}
				\end{fmfgraph*}\nonumber
			\end{align}
		\end{fmffile}
	\end{multicols}
	\caption{The two dominant tree level diagrams built out of the three vertices of the theory in the $t$-channel.}
	\label{fig:tree}
\end{figure}

Using the Feynman rules from the previous section, we first start with the two dominant tree level diagrams, which are drawn in \figref{fig:tree}. In terms of the Mandelstam variables, namely
\begin{align}
	s ~ &\coloneqq ~ - \left( p_1 + p_2 \right)^2 ~  = ~ - \left(p_3 + p_4 \right)^2 \, ,\\
	t ~ &\coloneqq ~ - \left( p_1 - p_3 \right)^2 ~ = ~ - \left( p_2 - p_4 \right)^2 \, ,\\
	u~ &\coloneqq ~ - \left( p_1 - p_4 \right)^2 ~ = ~ - \left( p_2 - p_3 \right)^2 \, , 
\end{align}
the two diagrams in the left and right panels of \figref{fig:tree} can be evaluated to find
\begin{align}
	i \mathcal{M}_{e^{-} e^{-}} ~ = ~ \dfrac{ - i q^2 s \left(\lambda - 2\right)}{2 \pi \lambda \left(\lambda - 2\right) - 2\pi R^{2} t \left(\lambda - 1\right)} \left[ 1 + \dfrac{t}{2s} - \dfrac{2 m^2}{s} - \dfrac{\lambda + 1}{s R^{2}} - \dfrac{1}{s R^{2}}\dfrac{\left(\lambda - 1\right)^{2}}{\lambda - 2}\right] \, 
\end{align}
and 
\begin{align}
	i \mathcal{M}_{e^{+} e^{-}} ~ = ~ \dfrac{ i q^2 s \left(\lambda - 2\right)}{2 \pi \lambda \left(\lambda - 2\right) - 2 \pi R^{2} t \left(\lambda - 1\right)} \left[ 1 + \dfrac{t}{2s} - \dfrac{2m^2}{s} - \dfrac{\lambda + 1}{sR^2} - \dfrac{1}{sR^2}\dfrac{\left(\lambda - 1\right)^{2}}{\lambda - 2} \right] \, ,
\end{align}
respectively. Here, we have made extensive use of the fact that the external particles are of course on-shell, namely
\begin{align}
p_1^2 ~ &= ~ - m^2 - \mu^2 \lambda \, , && p_3^2 ~ = ~ - m^2 - \mu^2 \, , \\
p_2^2 ~ &= ~ - m^2 - \mu^2 \, , && p_4^2 ~ = ~ - m^2 - \mu^2 \lambda \, .
\end{align}
We are primarily interested in the eikonal limit of scattering in this paper, which in flat space amounts to negligible momentum transfer $t\rightarrow 0$. Moreover, we demand the black hole eikonal condition that $M_{BH} E = M_{BH} \sqrt{s} \gg M^{2}_{Pl}$ which is equivalent to demanding that $s R^{2} \gg 1$ and that $s \gg m^{2}$. In this black hole eikonal limit, the above tree level amplitudes reduce to
\begin{align}
	i \mathcal{M}_{\text{tree}} ~ = ~ \pm \dfrac{ i q^2 s }{2 \pi \left(\ell^{2} + \ell + 1\right)} \, .
\end{align}
There is of course a third tree level diagram which is in the $s$-channel but it can be checked that this is of $\mathcal{O}\left(s^{0}\right)$ and therefore heavily sub-leading in the large $s$ limit. The above results were derived in the leading order near-horizon approximation of \secref{sec:leadingOrderApprox}. The analogous result in the shockwave approximation of \secref{sec:shockwaveApprox} is given by
\begin{align}
	i \mathcal{M}_{\text{tree}} ~ = ~ \pm \dfrac{ i q^2 s }{2 \pi \ell \left(\ell + 1\right)} \, .
\end{align}

\subsection{Loop diagrams and the eikonal ladder}
\begin{figure}[h!]
\vspace{1cm}
\centering
\begin{multicols}{3}
\begin{fmffile}{feyntwoLoop1}
\begin{fmfgraph*}(100,60)
\fmfstraight
\fmfleft{i1,i2}
\fmfright{o1,o2}
\fmflabel{$\hat{\varphi}^*_0$}{i1}
\fmflabel{$\hat{\varphi}$}{o1}
\fmflabel{$\hat{\varphi}^*$}{i2}
\fmflabel{$\hat{\varphi}_0$}{o2}
\fmf{dashes}{i1,v1}
\fmf{fermion}{v5,o1}
\fmf{fermion}{i2,v2}
\fmf{fermion}{v1,v3}
\fmf{dashes}{v3,v5}
\fmf{dashes}{v2,v4}
\fmf{fermion}{v4,v6}
\fmf{dashes}{v6,o2}
\fmf{photon, tension=0}{v1,v2}
\fmf{photon, tension=0}{v3,v4}
\fmf{photon, tension=0}{v5,v6}
\marrow{a}{up}{top}{$P$}{i2,o2}
\marrow{b}{down}{bot}{$P^\prime$}{i1,o1}
\end{fmfgraph*}
\end{fmffile}
    
\begin{fmffile}{feyntwoLoop2}
\begin{fmfgraph*}(100,60)
\fmfstraight
\fmfleft{i1,i2}
\fmfright{o1,o2}
\fmflabel{$\hat{\varphi}^*_0$}{i1}
\fmflabel{$\hat{\varphi}$}{o1}
\fmflabel{$\hat{\varphi}^*$}{i2}
\fmflabel{$\hat{\varphi}_0$}{o2}
\fmf{dashes}{i1,v1}
\fmf{fermion}{v5,o1}
\fmf{fermion}{i2,v2}
\fmf{fermion}{v1,v3}
\fmf{dashes}{v3,v5}
\fmf{dashes}{v2,v4}
\fmf{fermion}{v4,v6}
\fmf{dashes}{v6,o2}
\fmf{photon, tension=0}{v1,v2}
\fmf{photon, tension=0}{v3,v6}
\fmf{photon, tension=0}{v5,v4}
\marrow{a}{up}{top}{$P$}{i2,o2}
\marrow{b}{down}{bot}{$P^\prime$}{i1,o1}
\end{fmfgraph*}
\end{fmffile}
        
\begin{fmffile}{feyntwoLoop3}
\begin{fmfgraph*}(100,60)
\fmfstraight
\fmfleft{i1,i2}
\fmfright{o1,o2}
\fmflabel{$\hat{\varphi}^*_0$}{i1}
\fmflabel{$\hat{\varphi}$}{o1}
\fmflabel{$\hat{\varphi}^*$}{i2}
\fmflabel{$\hat{\varphi}_0$}{o2}
\fmf{dashes}{i1,v1}
\fmf{fermion}{v5,o1}
\fmf{fermion}{i2,v2}
\fmf{fermion}{v1,v3}
\fmf{dashes}{v3,v5}
\fmf{dashes}{v2,v4}
\fmf{fermion}{v4,v6}
\fmf{dashes}{v6,o2}
\fmf{photon, tension=0}{v1,v6}
\fmf{photon, tension=0}{v3,v2}
\fmf{photon, tension=0}{v5,v4}
\marrow{a}{up}{top}{$P$}{i2,o2}
\marrow{b}{down}{bot}{$P^\prime$}{i1,o1}
\end{fmfgraph*}
\end{fmffile}
\end{multicols}

\vspace{1.5cm}

\begin{multicols}{3}
\begin{fmffile}{feyntwoLoop4}
\begin{fmfgraph*}(100,60)
\fmfstraight
\fmfleft{i1,i2}
\fmfright{o1,o2}
\fmflabel{$\hat{\varphi}^*_0$}{i1}
\fmflabel{$\hat{\varphi}$}{o1}
\fmflabel{$\hat{\varphi}^*$}{i2}
\fmflabel{$\hat{\varphi}_0$}{o2}
\fmf{dashes}{i1,v1}
\fmf{fermion}{v5,o1}
\fmf{fermion}{i2,v2}
\fmf{fermion}{v1,v3}
\fmf{dashes}{v3,v5}
\fmf{dashes}{v2,v4}
\fmf{fermion}{v4,v6}
\fmf{dashes}{v6,o2}
\fmf{photon, tension=0}{v1,v4}
\fmf{photon, tension=0}{v3,v2}
\fmf{photon, tension=0}{v5,v6}
\marrow{a}{up}{top}{$P$}{i2,o2}
\marrow{b}{down}{bot}{$P^\prime$}{i1,o1}
\end{fmfgraph*}
\end{fmffile}
        
\begin{fmffile}{feyntwoLoop5}
\begin{fmfgraph*}(100,60)
\fmfstraight
\fmfleft{i1,i2}
\fmfright{o1,o2}
\fmflabel{$\hat{\varphi}^*_0$}{i1}
\fmflabel{$\hat{\varphi}$}{o1}
\fmflabel{$\hat{\varphi}^*$}{i2}
\fmflabel{$\hat{\varphi}_0$}{o2}
\fmf{dashes}{i1,v1}
\fmf{fermion}{v5,o1}
\fmf{fermion}{i2,v2}
\fmf{fermion}{v1,v3}
\fmf{dashes}{v3,v5}
\fmf{dashes}{v2,v4}
\fmf{fermion}{v4,v6}
\fmf{dashes}{v6,o2}
\fmf{photon, tension=0}{v1,v4}
\fmf{photon, tension=0}{v3,v6}
\fmf{photon, tension=0}{v5,v2}
\marrow{a}{up}{top}{$P$}{i2,o2}
\marrow{b}{down}{bot}{$P^\prime$}{i1,o1}
\end{fmfgraph*}
\end{fmffile}
    
\begin{fmffile}{feyntwoLoop6}
\begin{fmfgraph*}(100,60)
\fmfstraight
\fmfleft{i1,i2}
\fmfright{o1,o2}
\fmflabel{$\hat{\varphi}^*_0$}{i1}
\fmflabel{$\hat{\varphi}$}{o1}
\fmflabel{$\hat{\varphi}^*$}{i2}
\fmflabel{$\hat{\varphi}_0$}{o2}
\fmf{dashes}{i1,v1}
\fmf{fermion}{v5,o1}
\fmf{fermion}{i2,v2}
\fmf{fermion}{v1,v3}
\fmf{dashes}{v3,v5}
\fmf{dashes}{v2,v4}
\fmf{fermion}{v4,v6}
\fmf{dashes}{v6,o2}
\fmf{photon, tension=0}{v1,v6}
\fmf{photon, tension=0}{v3,v4}
\fmf{photon, tension=0}{v5,v2}
\marrow{a}{up}{top}{$P$}{i2,o2}
\marrow{b}{down}{bot}{$P^\prime$}{i1,o1}
\end{fmfgraph*}
\end{fmffile}
\end{multicols}
\vspace{0.15cm}
\caption{All leading $e^{-} e^{-}$ two loop diagrams in the black hole eikonal.}
\label{fig:twoloop}
\end{figure}
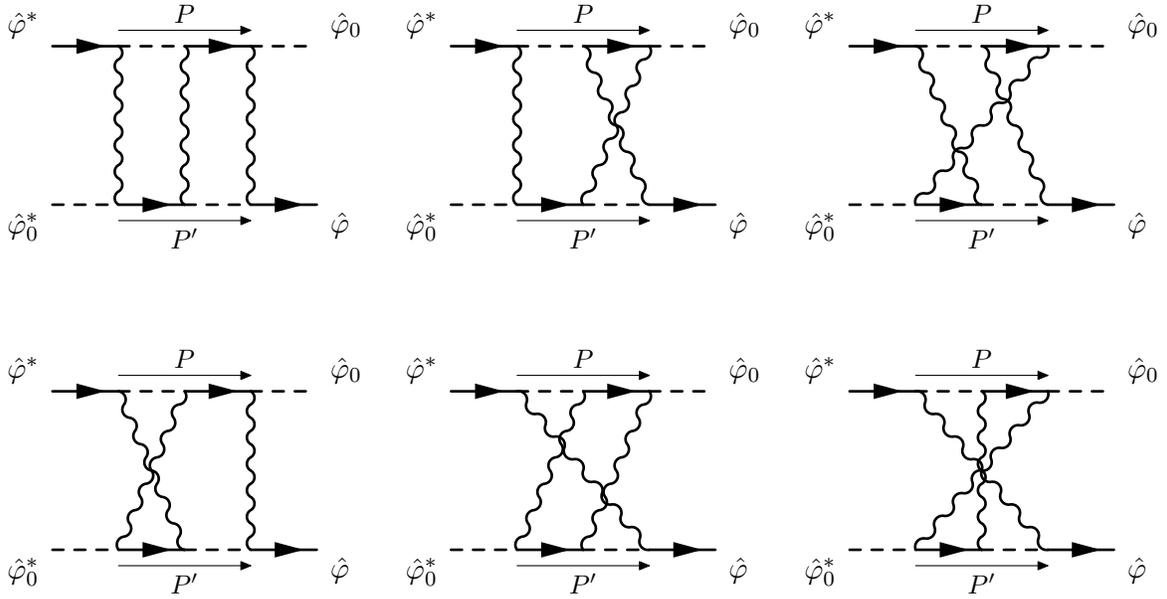

Loop diagrams in the black hole eikonal limit are dominated by the so-called ladder diagrams. The one and two loop diagrams are shown in \figref{fig:oneloop} and \figref{fig:twoloop}, respectively. 

\begin{figure}[h!]
\vspace{1cm}
\centering
\begin{multicols}{2}
\begin{fmffile}{feynoneLoop1}
\begin{fmfgraph*}(100,80)
\fmfstraight
\fmfleft{i1,i2}
\fmfright{o1,o2}
\fmflabel{$\hat{\varphi}_0^* (-p_2)$}{i1}
\fmflabel{$\hat{\varphi}_0 (p_4)$}{o1}
\fmflabel{$\hat{\varphi}^* (-p_1)$}{i2}
\fmflabel{$\hat{\varphi} (p_3)$}{o2}
\fmf{dashes}{i1,v1}
\fmf{dashes}{v3,o1}
\fmf{fermion}{i2,v2}
\fmf{fermion}{v1,v3}
\fmf{dashes}{v2,v4}
\fmf{fermion}{v4,o2}
\fmf{photon, tension=0}{v1,v2}
\fmf{photon, tension=0}{v3,v4}
\marrow{a}{up}{top}{$p_1$}{i2,v2}
\marrow{d}{right}{rt}{$k_1$}{v1,v2}
\marrow{g}{down}{bot}{$p_2 -k_1$}{v1,v3}
\marrow{h}{up}{top}{$p_1 +k_1$}{v2,v4}
\marrow{f}{right}{rt}{$k_2$}{v3,v4}
\marrow{e}{up}{top}{$p_3$}{v4,o2}
\marrow{c}{down}{bot}{$p_2$}{i1,v1}
\marrow{b}{down}{bot}{$p_4$}{v3,o1}
\end{fmfgraph*}
\end{fmffile}

\begin{fmffile}{feynoneLoop2}
\begin{fmfgraph*}(100,80)
\fmfstraight
\fmfleft{i1,i2}
\fmfright{o1,o2}
\fmflabel{$\hat{\varphi}_0^* (-p_2)$}{i1}
\fmflabel{$\hat{\varphi}_0 (p_4)$}{o1}
\fmflabel{$\hat{\varphi}^* (-p_1)$}{i2}
\fmflabel{$\hat{\varphi} (p_3)$}{o2}
\fmf{dashes}{i1,v1}
\fmf{dashes}{v3,o1}
\fmf{fermion}{i2,v2}
\fmf{fermion}{v1,v3}
\fmf{dashes}{v2,v4}
\fmf{fermion}{v4,o2}
\fmf{photon, tension=0}{v1,v4}
\fmf{photon, tension=0}{v2,v3}
\marrow{a}{up}{top}{$p_1$}{i2,v2}
\marrow{d}{left}{lft}{$k_1$}{v1,v2}
\marrow{g}{down}{bot}{$p_2 -k_2$}{v1,v3}
\marrow{h}{up}{top}{$p_1 +k_1$}{v2,v4}
\marrow{f}{right}{rt}{$k_2$}{v3,v4}
\marrow{e}{up}{top}{$p_3$}{v4,o2}
\marrow{c}{down}{bot}{$p_2$}{i1,v1}
\marrow{b}{down}{bot}{$p_4$}{v3,o1}
\end{fmfgraph*}
\end{fmffile}
\end{multicols}

\vspace{1.5cm}

\centering
\begin{multicols}{2}
\begin{fmffile}{feynoneLoop1ep}
\begin{fmfgraph*}(100,80)
\fmfstraight
\fmfleft{i1,i2}
\fmfright{o1,o2}
\fmflabel{$\hat{\varphi}_0 (-p_2)$}{i1}
\fmflabel{$\hat{\varphi}_0^* (p_4)$}{o1}
\fmflabel{$\hat{\varphi}^* (-p_1)$}{i2}
\fmflabel{$\hat{\varphi} (p_3)$}{o2}
\fmf{dashes}{v1,i1}
\fmf{dashes}{o1,v3}
\fmf{fermion}{i2,v2}
\fmf{fermion}{v3,v1}
\fmf{dashes}{v2,v4}
\fmf{fermion}{v4,o2}
\fmf{photon, tension=0}{v1,v2}
\fmf{photon, tension=0}{v3,v4}
\marrow{a}{up}{top}{$p_1$}{i2,v2}
\marrow{d}{right}{rt}{$k_1$}{v1,v2}
\marrow{g}{down}{bot}{$p_2 -k_1$}{v1,v3}
\marrow{h}{up}{top}{$p_1 +k_1$}{v2,v4}
\marrow{f}{right}{rt}{$k_2$}{v3,v4}
\marrow{e}{up}{top}{$p_3$}{v4,o2}
\marrow{c}{down}{bot}{$p_2$}{i1,v1}
\marrow{b}{down}{bot}{$p_4$}{v3,o1}
\end{fmfgraph*}\end{fmffile}

\begin{fmffile}{feynoneLoop2ep}
\begin{fmfgraph*}(100,80)
\fmfstraight
\fmfleft{i1,i2}
\fmfright{o1,o2}
\fmflabel{$\hat{\varphi}_0 (-p_2)$}{i1}
\fmflabel{$\hat{\varphi}_0^* (p_4)$}{o1}
\fmflabel{$\hat{\varphi}^* (-p_1)$}{i2}
\fmflabel{$\hat{\varphi} (p_3)$}{o2}
\fmf{dashes}{v1,i1}
\fmf{dashes}{o1,v3}
\fmf{fermion}{i2,v2}
\fmf{fermion}{v3,v1}
\fmf{dashes}{v2,v4}
\fmf{fermion}{v4,o2}
\fmf{photon, tension=0}{v1,v4}
\fmf{photon, tension=0}{v2,v3}
\marrow{a}{up}{top}{$p_1$}{i2,v2}
\marrow{d}{left}{lft}{$k_1$}{v1,v2}
\marrow{g}{down}{bot}{$p_2 -k_2$}{v1,v3}
\marrow{h}{up}{top}{$p_1 +k_1$}{v2,v4}
\marrow{f}{right}{rt}{$k_2$}{v3,v4}
\marrow{e}{up}{top}{$p_3$}{v4,o2}
\marrow{c}{down}{bot}{$p_2$}{i1,v1}
\marrow{b}{down}{bot}{$p_4$}{v3,o1}
\end{fmfgraph*}
\end{fmffile}
\end{multicols}
\vspace{0.15cm}
\caption{All leading one-loop diagrams in the black hole eikonal.}
\label{fig:oneloop}
\end{figure}
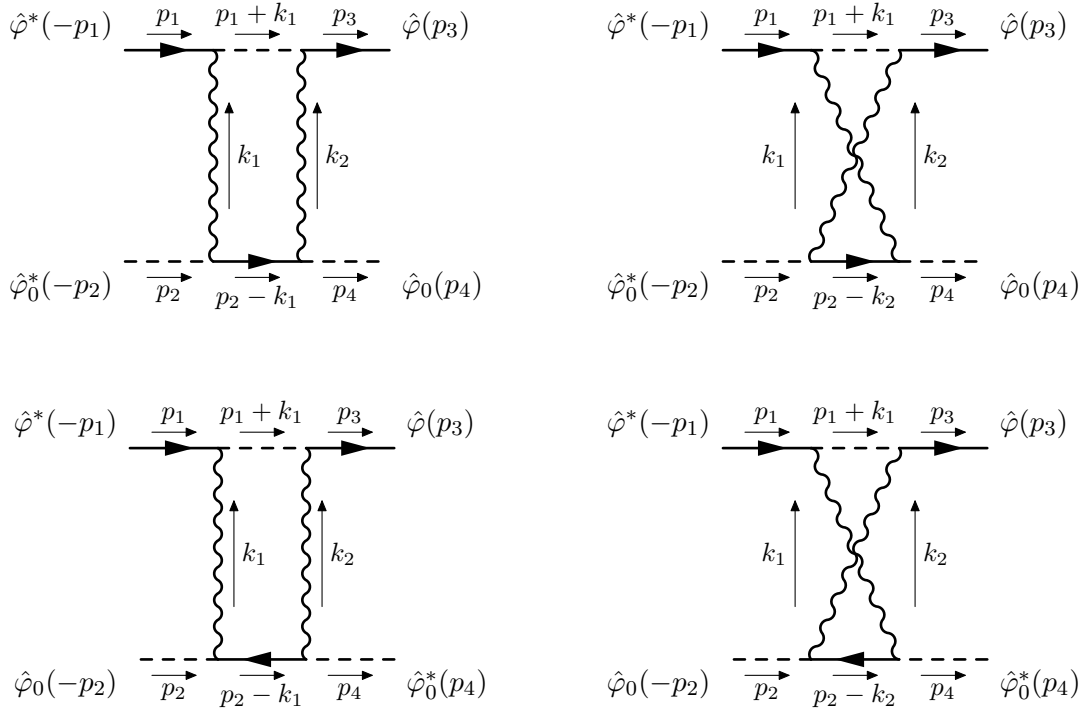

Following the analysis in the gravitational case \cite{Levy:1969cr, Kabat:1992tb, Gaddam:2020rxb, Gaddam:2020mwe}, a general loop diagram with $n$ virtual photons exchanged can be written as
\begin{align}\label{general loop amplitude 1}
i\mathcal{M}_{n} ~ &= ~ \left( i \dfrac{q}{\sqrt{4\pi} \, R}\right)^{2n} \int \prod_{j=1}^n \left[\dfrac{\mathrm{d}^2 k_j}{\left(2 \pi\right)^2} 4 p_a^1 p_b^2 \mathcal{P}^{ab}\left(k_j \right)\right] \times I \times \left(2 \pi\right)^2 \delta^{(2)} \left( \sum \nolimits_{j=1}^n k_j\right) \nonumber \\
&= ~ \left( i \dfrac{q}{\sqrt{4\pi} \, R}\right)^{2n} \left(\dfrac{s}{2}\right)^n \int \prod_{j=1}^n \left[\dfrac{\mathrm{d}^2 k_j}{\left(2 \pi\right)^2} 4 \mathcal{P}^{uv} \left(k_j \right) \right] \times I \times \left(2 \pi\right)^2 \delta^{(2)} \left( \sum \nolimits_{j=1}^n k_j \right) \, .
\end{align}
Of course, $n$ exchanged photons implies an $n-1$ loop amplitude. This equation is the two-dimensional analog of Eq. (3.1) in \cite{Levy:1969cr}, with electromagnetic vertices replacing the meson ones, and with $q = p_1 - p_3 = p_2 -p_4 = 0$. To get to the second equality, we assumed the two momenta to be light-like, i.e., $p_1 =\left(p_u^1 ,0\right)$, $p_2=\left(0,p_v^2\right)$. All the matter propagators to be inserted are contained in the quantity $I$, which can be derived analogously to \cite{Levy:1969cr}, resulting in
\begin{equation}
i \mathcal{M}_{n} ~ = ~ -\dfrac{q^2 s}{8 \pi R^{2} n!} \int \dfrac{\mathrm{d}^2 k}{\left(2 \pi \right)^2} 4 \mathcal{P}^{uv} \left(k\right) \int \mathrm{d}^2 x e^{- i k \cdot x} \left( i \chi \right)^{n-1} \, ,
\end{equation}
where the quantity $\chi$ has been defined as
\begin{multline}
\chi ~ \coloneqq ~ - \dfrac{i q^2 s}{8 \pi R^{2}} \int \dfrac{\mathrm{d}^2 k}{ \left(2 \pi\right)^2} 4 \mathcal{P}^{uv} \left( k \right) e^{- i k \cdot x} \times \left[ \dfrac{1}{- 2 p_1 \cdot k - i \epsilon} \dfrac{1}{2 p_2 \cdot k - i \epsilon} \right. \\
\left. \hspace{1cm} + \hspace{0.2mm} \dfrac{1}{- 2 p_1 \cdot k - i \epsilon} \dfrac{1}{ - 2 p_2 \cdot k - i \epsilon} + \dfrac{1}{2 p_1 \cdot k - i \epsilon}\dfrac{1}{2 p_2 \cdot k - i \epsilon} \right. \\
\left. \hspace{1.5cm}+\hspace{0.2mm} \dfrac{1}{2 p_1 \cdot k - i \epsilon} \dfrac{1}{ - 2 p_2 \cdot k - i \epsilon}\right] \, .
\end{multline}
The expression in square brackets can be rewritten in a more convenient form as
\begin{multline}
\chi ~ = ~ - \dfrac{ i q^2 s}{8 \pi R^{2}} \int \dfrac{\mathrm{d}^2 k}{\left( 2 \pi\right)^2} 4 \mathcal{P}^{uv} \left(k\right) e^{ - i k \cdot x} \left( \dfrac{1}{2 p_1 \cdot k + i \epsilon} - \dfrac{1}{2 p_1 \cdot k - i \epsilon} \right) \\
\times \left( \dfrac{1}{2 p_2 \cdot k + i \epsilon} - \dfrac{1}{2 p_2 \cdot k - i \epsilon}\right) \, .
\end{multline}
Now, making use of the identity
\begin{equation}
\dfrac{1}{ x + i \epsilon} - \dfrac{1}{x - i \epsilon} ~ = ~ - 2 \pi i \delta\left(x\right) \, , 
\end{equation}
we arrive at a simple expression for $\chi$:
\begin{align}\label{chi}
\chi ~ &= ~ - \dfrac{ i q^2 s}{8 \pi R^{2}} \int \dfrac{\mathrm{d}^2 k}{\left(2 \pi\right)^2} 4 \mathcal{P}^{uv} \left(k\right) e^{ - i k \cdot x} \left( - 2 \pi i \right)^2 \delta\left( 2 p_1 \cdot k\right) \delta\left( 2 p_2 \cdot k\right) \nonumber \\
&= ~ - \dfrac{q^2}{ 4 \pi R^{2}} \mathcal{P}^{uv} \left(0\right) \nonumber \\
&= ~ 
    \begin{cases}
      - \dfrac{q^2}{ 4 \pi \left(\lambda - 1 \right)} & \text{in the shockwave approximation of \secref{sec:shockwaveApprox}} \, ,\\
      - \dfrac{q^2}{4 \pi \lambda} & \text{in the leading order approximation of \secref{sec:leadingOrderApprox}} \, .
    \end{cases}
\end{align}
Since the resulting eikonal function, conveniently enough, does not depend on spacetime coordinates, we may write
\begin{align}
i \mathcal{M}_{n} ~ &= ~ - \dfrac{q^2 s}{8 \pi R^{2} n!} \left( i \chi\right)^{n-1} \int \dfrac{\mathrm{d}^2 k}{\left(2 \pi \right)^2} 4 \mathcal{P}^{uv} \left(k\right) \int \mathrm{d}^2 x e^{- i k \cdot x} \nonumber \\
&= ~ - \dfrac{q^2 s}{8 \pi R^{2} n!} \left( i \chi\right)^{n-1} \int \dfrac{\mathrm{d}^2 k}{\left( 2 \pi \right)^2} 4 \mathcal{P}^{uv}\left(k\right) \left(2 \pi\right)^2 \delta^{(2)} \left( k \right) \nonumber \\
&= ~ 2 s \dfrac{\left( i \chi \right)^n}{n!} \, .
\end{align}
The complete resummed perturbatively exact amplitude is therefore given by
\begin{align}\label{general loop amplitude pre-final}
i \mathcal{M} ~ &= ~ i \sum_{n} \mathcal{M}_{n} ~ = ~ 2 s \left[\exp(i \chi) - 1\right] \, .
\end{align}
Inserting \eqref{chi} in \eqref{general loop amplitude pre-final} in the above equation, and recalling that $\lambda = \ell^2+\ell+1$, we find
\begin{equation}\label{final expression general loop amplitude pp case}
i \mathcal{M} ~ = ~ \begin{cases}
      4 p_{\text{in}} p_{\text{out}} \left[ \exp \left( - \dfrac{i}{4 \pi} \dfrac{q^2}{\ell^2 + \ell}\right) - 1\right] & \text{shockwave approximation} \, ,\vspace{2mm}\\
      4 p_{\text{in}} p_{\text{out}} \left[\exp \left( - \dfrac{i}{4 \pi} \dfrac{q^2}{\ell^2 + \ell + 1}\right) - 1\right] & \text{leading order approximation} \, ,
    \end{cases}
\end{equation}
where we also relabelled the external momenta as $p_{in}$ and $p_{out}$. This amplitude is a result of diagrams of the $e^{-} e^{-}$ kind in \figref{fig:twoloop}. Considering the remaining case of $e^{-} e^{+}$ scattering results in an overall sign in the phase of the exponent. These two cases can be combined into a single formula, resulting in
\begin{equation}\label{final expression general loop amplitude}
i \mathcal{M} ~ = ~ \begin{cases}
      4 p_{\text{in}} p_{\text{out}} \left[ \exp \left( - \dfrac{i}{4 \pi} \dfrac{q_{\text{in}}q_{\text{out}}}{\ell^2 + \ell}\right) - 1\right] & \text{shockwave approximation} \, ,\vspace{2mm}\\
      4 p_{\text{in}} p_{\text{out}} \left[\exp \left( - \dfrac{i}{4 \pi} \dfrac{q_{\text{in}}q_{\text{out}}}{\ell^2 + \ell + 1}\right) - 1\right] & \text{leading order approximation} \, ,
    \end{cases}
\end{equation}
where $q_{\text{in}}$ and $q_{\text{out}}$ are the asymptotic charges of the in-particle and out-particle, respectively. For particles, we have that $q_{\text{in}/\text{out}}=-q$ and that $q_{\text{in}/\text{out}}=q$ for antiparticles. The relation between the scattering amplitude and the S-matrix is given by
\begin{equation}\label{relation S-M}
\left\langle \text{out} | S-\mathds{1} | \text{in} \right\rangle ~ = ~ (2\pi)^2 \delta^{(2)}\left(p_1 +p_2 -p_3 -p_4 \right)i\left\langle \text{out} | \mathcal{M} | \text{in} \right\rangle \, .
\end{equation}
For instance, considering particles, the in- and out-states can be defined as
\begin{align}
|\text{in}\rangle ~ &\coloneqq ~ |\text{in}_{1}\rangle \otimes |\text{in}_{2}\rangle ~ = ~ \frac{1}{2}\left(a_{00}^{\dagger}(p_1)+a_{\ell m}^{\dagger}(p_1) \right)\left(a_{00}^{\dagger}(p_2)+a_{\ell m}^{\dagger}(p_2) \right)|0\rangle  \, , \\
|\text{out}\rangle ~ &\coloneqq ~ |\text{out}_{1}\rangle \otimes |\text{out}_{2}\rangle ~ = ~ \frac{1}{2}\left(a_{00}^{\dagger}(p_3)+a_{\ell m}^{\dagger}(p_3) \right)\left(a_{00}^{\dagger}(p_4)+a_{\ell m}^{\dagger}(p_4) \right)|0\rangle \, .
\end{align}
A similar definition exists for antiparticles. In the free theory, a straightforward calculation leads to\footnote{In comparison to \cite{Gaddam:2020rxb, Gaddam:2020mwe}, the overall kinematic factor differs by a factor of 2 owing to a different pre-factor in the definition of the Mandelstam variables.}
\begin{equation}
\left\langle \text{out} | \text{in} \right\rangle ~ = ~ 2s(2\pi)^2 \left[\delta(p_1-p_3)\delta(p_2 -p_4 ) + \delta(p_1-p_4)\delta(p_2 -p_3 )\right] \, .
\end{equation}
On the other hand, using Eq. (\ref{general loop amplitude pre-final}), we may write the interacting piece as
\begin{equation}
\left\langle \text{out} | S-\mathds{1} | \text{in} \right\rangle ~ = ~ 2s(2\pi)^2 (e^{i\chi}-1) \left[\delta(p_1 -p_3)\delta(p_2 -p_4 ) + \delta(p_1 -p_4)\delta(p_2 -p_3 )\right] \, . 
\end{equation}
Putting it all together, Eq. (\ref{relation S-M}) in the operator notation gives
\begin{equation}
S ~ = ~ \mathds{1} e^{i\chi} ~ = ~ \begin{cases}
      \mathds{1} \exp \left( - \dfrac{i}{4 \pi} \dfrac{q_{\text{in}}q_{\text{out}}}{\ell^2 + \ell}\right) & \text{shockwave approximation} \, ,\vspace{2mm}\\
      \mathds{1}\exp \left( - \dfrac{i}{4 \pi} \dfrac{q_{\text{in}}q_{\text{out}}}{\ell^2 + \ell + 1}\right) & \text{leading order approximation} \, .
    \end{cases}
\end{equation}
 This result agrees with the expectation from the first quantised shockwave S-matrix in \eqref{eqn:shockwaveQMSmatrix} up to a curious factor of $4\pi$, which we address in \appref{app:4pi}.

\subsection{One-loop diagrams with the four vertex}\label{app:seagull}
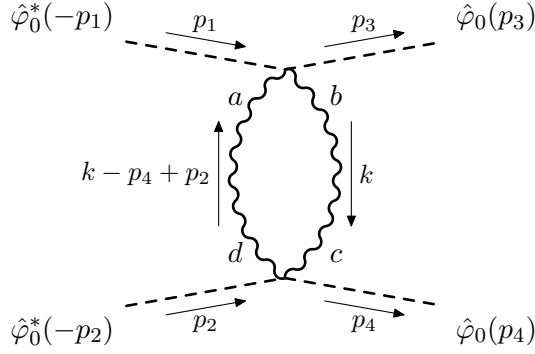
\begin{figure}[h!]
\centering
\begin{fmffile}{feyn1loop4vertex}
\begin{align*}
\begin{fmfgraph*}(120,100)
\fmfstraight
\fmfleft{i1,i2}
\fmfright{o1,o2}
\fmflabel{$\hat{\varphi}_0^* (-p_2)$}{i1}
\fmfv{label=$\hat{\varphi}_0(p_4)$}{o1}
\fmflabel{$\hat{\varphi}_0 (p_3)$}{o2}
\fmflabel{$\hat{\varphi}_0^* (-p_1)$}{i2}
\fmf{dashes,tension=1.5}{i1,v1}
\fmf{dashes,tension=1.5}{v1,o1}
\fmf{photon,left=0.5,tension=0.2}{v1,v2}
\fmf{photon,left=0.5,tension=0.2}{v2,v1}
\fmf{dashes,tension=1.5}{i2,v2}
\fmf{dashes,tension=1.5}{v2,o2}
\marrow{a}{up}{top}{$p_1$}{i2,v2}
\marrow{c}{down}{bot}{$p_2$}{i1,v1}
\marrow{b}{up}{top}{$p_3$}{v2,o2}
\marrow{d}{down}{bot}{$p_4$}{v1,o1}
\Marrow{e}{right}{rt}{$k$}{v2,v1}{25}
\Marrow{f}{left}{lft}{$k-p_4+p_2$}{v1,v2}{25}
\fmfv{label=$a\qquad \quad b$,label.angle=-90}{v2}
\fmfv{label=$d\qquad \quad c$,label.angle=90}{v1}
\end{fmfgraph*}
\end{align*}
\end{fmffile}
\caption{One-loop diagram arising from the four vertex involving the even mode.}
\label{fig:diagram four-vertex 1}
\end{figure}

In the second quantised theory, there are further corrections to the eikonal amplitudes that are not visible in the first quantised formalism of 't Hooft \cite{tHooft:1996rdg}. In the present case, the first such correction arises from the four vertex. The contributions of the four vertex at tree-level are naturally sub-leading in comparison to those of the three-vertex. This is down to the simple fact that the vertex does not contain momenta. Therefore, in the limit of large energies of centre of mass, the three-vertex naturally dominates. As it turns out, this is also true at loop level as we will demonstrate in this appendix. In what follows, we consider one-loop diagrams of the type drawn in \figref{fig:diagram four-vertex 1}.
\vspace{0.4cm}

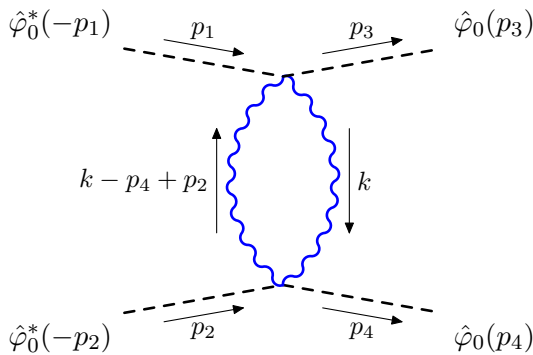
\begin{figure}[ht]
\centering
\begin{fmffile}{feyn1loop4vertexodd}
\begin{align*}
\begin{fmfgraph*}(120,100)
\fmfstraight
\fmfleft{i1,i2}
\fmfright{o1,o2}
\fmflabel{$\hat{\varphi}_0^* (-p_2)$}{i1}
\fmfv{label=$\hat{\varphi}_0(p_4)$}{o1}
\fmflabel{$\hat{\varphi}_0 (p_3)$}{o2}
\fmflabel{$\hat{\varphi}_0^* (-p_1)$}{i2}
\fmf{dashes,tension=1.5}{i1,v1}
\fmf{dashes,tension=1.5}{v1,o1}
\fmf{photon,foreground=blue,left=0.5,tension=0.2}{v1,v2}
\fmf{photon,foreground=blue,left=0.5,tension=0.2}{v2,v1}
\fmf{dashes,tension=1.5}{i2,v2}
\fmf{dashes,tension=1.5}{v2,o2}
\marrow{a}{up}{top}{$p_1$}{i2,v2}
\marrow{c}{down}{bot}{$p_2$}{i1,v1}
\marrow{b}{up}{top}{$p_3$}{v2,o2}
\marrow{d}{down}{bot}{$p_4$}{v1,o1}
\Marrow{e}{right}{rt}{$k$}{v2,v1}{25}
\Marrow{f}{left}{lft}{$k-p_4+p_2$}{v1,v2}{25}
\end{fmfgraph*}
\end{align*}
\end{fmffile}
\caption{One-loop four-vertex diagram involving the odd mode of the photon.}
\label{fig:diagram four-vertex 2}
\end{figure}

Using the Feynman rules presented in \secref{sec:FeynRules}, we write the amplitude as follows\footnote{Note that momentum conservation implies that $k-p_4 +p_2 -k+p_3 -p_1=0$.}:
\begin{multline}\label{seagull}
i\mathcal{M}_{\text{seagull}}^{\text{even}}=\int\frac{\mathrm{d}^2 k}{\left(2\pi\right)^2}\left[\frac{-i}{k^2 +\Tilde{m}^2 -i\epsilon}\left(\eta^{bc}+\frac{R^2k^b k^c}{\lambda -1}\right)\left(-\frac{2iq^2}{4\pi R^2} \eta_{ab}\right)\right.\\
\left. \times\hspace{0.5mm}\frac{-i}{k^{\prime 2} +\Tilde{m}^2 -i\epsilon}\left(\eta^{da}+\frac{R^2 k^d k^a}{\lambda -1}\right)\left(-\frac{2iq^2}{4\pi R^2} \eta_{cd}\right)\right],
\end{multline}
with $\Tilde{m}^2 \coloneqq \frac{\lambda -1}{R^2}$, $k^\prime \coloneqq k-\Tilde{p}$, $\Tilde{p}\coloneqq p_4 -p_2$. It is easy to see that Eq. (\ref{seagull}) can be split into four contributions. We can thus write 
\begin{equation}
i\mathcal{M}_{\text{seagull}}^{\text{even}}=I_1 +I_2 +I_3 +I_4,
\end{equation}
where the following quantities have been defined: 
\begin{align}
I_1 &\coloneqq \frac{q^4}{2\pi^2 R^4}\int\frac{\mathrm{d}^2 k}{\left(2\pi\right)^2}\frac{1}{\left(k^2 +\Tilde{m}^2 -i\epsilon \right)\left(k^{\prime^2} +\Tilde{m}^2 -i\epsilon \right)},\label{I_1}\\
I_2 &\coloneqq \frac{q^4}{4\pi^2 R^2}\frac{1}{\lambda -1}\int\frac{\mathrm{d}^2 k}{\left(2\pi\right)^2}\frac{k^{\prime 2}}{\left(k^2 +\Tilde{m}^2 -i\epsilon \right)\left(k^{\prime^2} +\Tilde{m}^2 -i\epsilon \right)},\label{I_2}\\
I_3 &\coloneqq \frac{q^4}{4\pi^2 R^2}\frac{1}{\lambda -1}\int\frac{\mathrm{d}^2 k}{\left(2\pi\right)^2}\frac{k^{\prime 2}}{\left(k^2 +\Tilde{m}^2 -i\epsilon \right)\left(k^{\prime 2} +\Tilde{m}^2 -i\epsilon \right)},\label{I_3}\\
I_4 &\coloneqq \frac{q^4}{4\pi^2}\frac{1}{\left(\lambda-1\right)^2}\int\frac{\mathrm{d}^2 k}{\left(2\pi\right)^2}\frac{\eta_{ab}\eta_{cd}k^b k^c k^{\prime d}k^{\prime a}}{\left(k^2 +\Tilde{m}^2 -i\epsilon \right)\left(k^{\prime 2} +\Tilde{m}^2 -i\epsilon \right)}.\label{I_4}
\end{align}
In what follows, we will work in coordinates where the near-horizon two-dimensional flat metric is diagonal, instead of the light-cone variants we have used so far. These two sets of coordinates are related by
\begin{equation}
u = \dfrac{1}{\sqrt{2}}\left(x^0 +x^1 \right) \, , \hspace{0.2cm} v = \dfrac{1}{\sqrt{2}}\left(x^0 -x^1\right) \, .  
\end{equation}
We will employ dimensional regularization and start by considering \eqref{I_1}, temporarily suppressing the $i\epsilon$'s for notational convenience, where we shift to $d$ dimensions:
\begin{equation}\label{integral in 1}
\int\frac{\mathrm{d}^2 k}{\left(2\pi\right)^2}\frac{1}{\left(k^2 +\Tilde{m}^2 \right)\left(k^{\prime^2} +\Tilde{m}^2\right)}\hspace{0.2cm}\to \hspace{0.2cm}\int\frac{\mathrm{d}^d k}{\left(2\pi\right)^d}\frac{1}{\left(k^2 +\Tilde{m}^2 \right)\left(k^{\prime^2} +\Tilde{m}^2\right)}, \hspace{0.2cm}d=2+\varepsilon .
\end{equation}
Using the familiar Feynman trick 
\begin{equation}
\dfrac{1}{AB} = \int_0^1 \mathrm{d}x \dfrac{1}{\left[A + ( B - A ) x \right]^2} \, ,    
\end{equation}
the above integral can be written as
\begin{equation}\label{before expanding}
\int\frac{\mathrm{d}^d k}{\left(2\pi\right)^d}\frac{1}{\left(k^2 +\Tilde{m}^2 \right)\left(k^{\prime^2} +\Tilde{m}^2\right)}=\int_0^1 \mathrm{d}x\int \frac{\mathrm{d}^d k}{(2\pi)^d}\frac{1}{\left[(k-xq)^2 +\Tilde{p}^2 x(1-x)+\Tilde{m}^2 \right]^2} \, .   
\end{equation}
Shifting the $k$ integral above by $k\to k+x\Tilde{p}$ and performing a Wick rotation (we substitute $k^0 =ik_E^0$) leads to
\begin{equation}\label{delta definition}
\int\frac{\mathrm{d}^d k}{\left(2\pi\right)^d}\frac{1}{\left(k^2 +\Tilde{m}^2 \right)\left(k^{\prime^2} +\Tilde{m}^2\right)} = i\int_0^1 \mathrm{d}x \int \frac{\mathrm{d}^d k_E}{(2\pi)^d}\frac{1}{\left(k_E^2 +\Delta \right)^2} \, ,
\end{equation}
where we defined $\Delta\coloneqq \Tilde{p}^2 x(1-x)+\Tilde{m}^2$. Momentum integrals of the kind above can be expressed in terms of gamma functions
\begin{equation}\label{wick rotation result}
\int\frac{\mathrm{d}^d k_E}{(2\pi)^d}\frac{1}{\left(k_E^2 +\Delta \right)^\alpha}=\frac{1}{(4\pi)^{\frac{d}{2}}} \frac{\Gamma\left(\alpha-\frac{d}{2}\right)}{\Gamma(\alpha)}\Delta^{\frac{d}{2}-\alpha} \, . 
\end{equation}
In our case, with $\alpha =2$ and $d = 2 + \varepsilon$, we have
\begin{equation}
\int\frac{\mathrm{d}^d k}{\left(2\pi\right)^d}\frac{1}{\left(k^2 +\Tilde{m}^2 \right)\left(k^{\prime^2} +\Tilde{m}^2\right)}=\frac{i}{4\pi}(4\pi)^{-\frac{\epsilon}{2}}\Gamma\left(1-\frac{\varepsilon}{2}\right)M^{\varepsilon-2}\int_0^1 \mathrm{d}x \left(\frac{\Delta}{M^2}\right)^{\frac{\varepsilon}{2}-1} \, ,    
\end{equation}
where we introduced an auxiliary mass parameter, $M$. This allows us to consider small-$\varepsilon$ expansions of the following dimensionless quantities:
\begin{align}
(4\pi)^{-\frac{\varepsilon}{2}}&=1-\frac{\varepsilon}{2}\ln (4\pi)+\ldots ,\\
\Gamma\left(1-\frac{\varepsilon}{2}\right)&=-\frac{\varepsilon}{2}\Gamma\left(-\frac{\varepsilon}{2}\right)=-\frac{\varepsilon}{2}\left(-\frac{2}{\varepsilon}-\gamma_E +\ldots\right),\\
\left(\frac{\Delta}{M^2}\right)^{\frac{\varepsilon}{2}-1}&=\frac{M^2}{\Delta}+\frac{\varepsilon}{2}\frac{M^2 \ln\left(\Delta/M^2 \right)}{\Delta}+\ldots,
\end{align}
where $\gamma_E \approx 0.5772$ is the Euler-Mascheroni constant. Using these expressions we obtain\footnote{Expanding before performing the integral is allowed in this case since each term in the expansion, when integrated, converges.} 
\begin{equation}
\int\frac{\mathrm{d}^d k}{\left(2\pi\right)^d}\frac{1}{\left(k^2 +\Tilde{m}^2 \right)\left(k^{\prime^2} +\Tilde{m}^2\right)}=\frac{i}{4\pi}\int_0^1 \mathrm{d}x\frac{1}{\Delta}=\frac{iM^{\varepsilon}}{4\pi}\int_0^1 \mathrm{d}x\frac{1}{\Tilde{p}^2 x(1-x)+\Tilde{m}^2}.
\end{equation}
In principle, various cases must be considered, depending on the values $\Tilde{p}^2$ can assume; however, since we are only interested in the limit $\Tilde{p}\to 0$ (negligible momentum transfer), we directly expand the integrand and consider the first term of such an expansion\footnote{Again, this is allowed because each term in the expansion, when integrated, converges.}. We have:
\begin{equation}\label{expansion in p}
\frac{1}{\Tilde{p}^2 x(1-x)+\Tilde{m}^2}=\frac{1}{\Tilde{m}^2}+\frac{\Tilde{p}^2 (x-1)x}{\Tilde{m}^4}+\frac{\Tilde{p}^4 (x-1)^2 x^2}{\Tilde{m}^6}+\mathcal{O}(\Tilde{p}^6) \, .
\end{equation}
Therefore, in this specific limit the result of the above integral is
\begin{equation}\label{finite contribution}
\int\frac{\mathrm{d}^d k}{\left(2\pi\right)^d}\frac{1}{\left(k^2 +\Tilde{m}^2 \right)\left(k^{\prime^2} +\Tilde{m}^2\right)}=\frac{iM^{\varepsilon}}{4\pi\tilde{m}^2}+\mathcal{O}(\varepsilon).
\end{equation}
Before proceeding, let us make another observation about dimensions. When calculating Feynman diagrams in $2+\varepsilon$ spacetime dimensions, the coupling constants will carry the dimension that is appropriate for the theory in $2+\varepsilon$ dimensions. For the scalar quantum electrodynamics built here, the dimension of the effective coupling constant turns out to be equal to $1-\varepsilon/2$ in mass units. On the other hand, in the 2-dimensional case we have that $[\mu q]=1$ (of course, integrating the sphere out does not change the dimensions of the quantity $q$). Therefore, in order to ensure that dimensional counting remains consistent throughout the calculations, we make again use of the auxiliary parameter $M$ and write the effective coupling constant as $M^{-\varepsilon/2}\mu q$. Putting it all together (taking into account the various prefactors), we now write down the final expression for $I_1$ in $2+\varepsilon$ spacetime dimensions, in the limit $\Tilde{p}\to 0$:
\begin{equation}
I_1^d \bigr\rvert_{\Tilde{p}\to 0}=\frac{iM^{-\varepsilon}q^4}{8\pi^3 R^2}\frac{1}{\lambda-1}+\mathcal{O}(\varepsilon).
\end{equation}
Let us now consider the second contribution, namely $I_2$. Ignoring the prefactors for a moment, shifting to $d$ dimensions, and writing $k^\prime =k^\prime +\Tilde{m}^2 -\Tilde{m}^2$, leads to
\begin{equation}\label{second piece}
\int\frac{\mathrm{d}^d k}{\left(2\pi\right)^d}\frac{k^{\prime 2}}{\left(k^2 +\Tilde{m}^2 \right)\left(k^{\prime^2} +\Tilde{m}^2\right)}=\int\frac{\mathrm{d}^d k}{\left(2\pi\right)^d}\frac{1}{k^2 +\Tilde{m}^2}- \int\frac{\mathrm{d}^d k}{\left(2\pi\right)^d}\frac{\Tilde{m}^2}{\left(k^2 +\Tilde{m}^2 \right)\left(k^{\prime 2}+\Tilde{m}^2 \right)}.
\end{equation}
The first term can be quite easily computed by performing a Wick rotation to use (\ref{wick rotation result}) with $\alpha=1$, where the role of $\Delta$ is now played by $\Tilde{m}^2$. We have:
\begin{equation}\label{special case alpha=1}
\int\frac{\mathrm{d}^d k}{\left(2\pi\right)^d}\frac{1}{k^2 +\Tilde{m}^2}=\frac{i}{(4\pi)^{\frac{d}{2}}}\Gamma\left(1-\frac{d}{2}\right)\left(\Tilde{m}^2 \right)^{\frac{d}{2}-1}. 
\end{equation}
We now substitute $d=2+\varepsilon$ and expand in powers of $\varepsilon$, keeping track of possible poles at $\varepsilon=0$. In terms of $\varepsilon$, Eq. (\ref{special case alpha=1}) then becomes
\begin{equation}\label{final second integral first piece}
\int\frac{\mathrm{d}^d k}{\left(2\pi\right)^d}\frac{1}{k^2 +\Tilde{m}^2}=\frac{i}{4\pi}\left(4\pi \right)^{-\frac{\varepsilon}{2}}\Gamma\left(-\frac{\varepsilon}{2}\right)\left(\Tilde{m}^2 \right)^{\frac{\varepsilon}{2}}. 
\end{equation}
Introducing $M$ as before and rearranging, we get
\begin{equation}
\int\frac{\mathrm{d}^d k}{\left(2\pi\right)^d}\frac{1}{k^2 +\Tilde{m}^2}=\frac{iM^{\varepsilon}}{4\pi}\left(4\pi \right)^{-\frac{\varepsilon}{2}}\Gamma\left(-\frac{\varepsilon}{2}\right)\left(\frac{\Tilde{m}^2}{M^2}\right)^{\frac{\varepsilon}{2}}. 
\end{equation}
We can now safely expand, obtaining
\begin{equation}
\int\frac{\mathrm{d}^d k}{\left(2\pi\right)^d}\frac{1}{k^2 +\Tilde{m}^2}=-\frac{iM^{\varepsilon}}{2\pi} \left[\frac{1}{\varepsilon}+\frac{1}{2}\gamma_E +\frac{1}{2}\ln\left(\frac{1}{4\pi}\frac{\Tilde{m}^2}{M^2}\right)+\mathcal{O}(\varepsilon)\right],
\end{equation}
which is dimensionally consistent. Concerning the second term in (\ref{second piece}), it has already been computed. Putting it all together, we obtain the final result for $I_2$:
\begin{equation}
I_2^d \bigr\rvert_{\Tilde{p}\to 0}=-\frac{iM^{-\varepsilon}q^4}{8\pi^3 R^2}\frac{1}{\lambda-1}\left[\frac{1}{\varepsilon}+\frac{1}{2}\left(\gamma_E +1 \right)+\frac{1}{2}\ln\left(\frac{\lambda-1}{4\pi R^2 M^2}\right)+\mathcal{O}(\varepsilon)\right].
\end{equation}
Now, looking at the third contribution to the amplitude, Eq. (\ref{I_3}), we notice that it is equal to Eq. (\ref{I_2}) upon shifting the momentum $k$, $k \to k+\Tilde{p}$. Therefore, we move on to the fourth contribution, $I_4$. Let us first consider the numerator of the integrand. By recalling how $k^\prime$ is defined, it can be split as
\begin{align}
\eta_{ab}\eta_{cd}k^b k^c k^{\prime d}k^{\prime a} ~ &= ~ k_a k^{\prime a}k_c k^{\prime c} ~ = ~ (k\cdot k^\prime)^2 \nonumber\\
&= ~ k^2 k^{\prime 2}-k^{\prime 2}(\Tilde{p}\cdot k)+k^2 (\Tilde{p}\cdot k^\prime)-(\Tilde{p}\cdot k)(\Tilde{p}\cdot k^\prime).
\end{align}
Thus, shifting to $2+\varepsilon$ spacetime dimensions, the integral in \eqref{I_4} can be written as
\begin{equation}\label{splitting fourth integral}
\int\frac{\mathrm{d}^d k}{\left(2\pi\right)^d}\frac{k^2 k^{\prime 2}-k^{\prime 2}(\Tilde{p}\cdot k)+k^2 (\Tilde{p}\cdot k^\prime)-(\Tilde{p}\cdot k)(\Tilde{p}\cdot k^\prime)}{\left(k^2 +\Tilde{m}^2 \right)\left(k^{\prime 2} +\Tilde{m}^2 \right)}.
\end{equation}
As we can see, $I_4$ has been split into four contributions. The first gives
\begin{equation}\label{intermediate 4-1}
\int \frac{\mathrm{d}^d k}{(2\pi)^d}\frac{k^2 k^{\prime 2}}{\left(k^2 +\Tilde{m}^2 \right)\left(k^{\prime 2} +\Tilde{m}^2 \right)}=\int \frac{\mathrm{d}^d k}{(2\pi)^d}\frac{k^{\prime 2}}{k^{\prime 2}+\Tilde{m}^2}- \int \frac{\mathrm{d}^d k}{(2\pi)^d}\frac{\Tilde{m}^2 k^{\prime 2}}{\left(k^2 +\Tilde{m}^2 \right)\left(k^{\prime 2} +\Tilde{m}^2 \right)}.   
\end{equation}
The second piece of the right-hand side of the above equation has already been calculated before. The first one can be easily computed by Wick rotating and making use of the identity
\begin{equation}
\int\frac{d^d k_E}{(2\pi)^d}\frac{k_E^2}{\left(k_E^2 +\Delta \right)^\alpha}=\frac{d}{2}\frac{1}{(4\pi)^{\frac{d}{2}}}\frac{\Gamma\left(\alpha-\frac{d}{2}-1\right)}{\Gamma(\alpha)}\Delta^{\frac{d}{2}-\alpha +1} \, .  
\end{equation}
In terms of $\varepsilon$, by setting $\alpha=1$ and shifting $k\to k+\Tilde{p}$, we can write
\begin{equation}
\int \frac{\mathrm{d}^d k}{(2\pi)^d}\frac{k^{\prime 2}}{k^{\prime 2}+\Tilde{m}^2}=\frac{i\Tilde{m}^2}{4\pi}(1+\varepsilon)(4\pi)^{-\frac{\varepsilon}{2}}\Gamma\left(-1-\frac{\varepsilon}{2}\right)\left(\Tilde{m}^2 \right)^{\frac{\varepsilon}{2}}. 
\end{equation}
Moreover, inserting $M$ and expanding in powers of $\varepsilon$, we end up with
\begin{equation}
\int \frac{\mathrm{d}^d k}{(2\pi)^d}\frac{k^{\prime 2}}{k^{\prime 2}+\Tilde{m}^2}=\frac{iM^{\varepsilon}\Tilde{m}^2}{2\pi}\left[\frac{1}{\varepsilon}+\frac{1}{2}\left(\gamma_E +1\right)+\frac{1}{2}\ln\left(\frac{1}{4\pi}\frac{\Tilde{m}^2}{M^2}\right)+\mathcal{O}(\varepsilon)\right].    
\end{equation}
Therefore, Eq. (\ref{intermediate 4-1}) results in
\begin{equation}
\int \frac{\mathrm{d}^d k}{(2\pi)^d}\frac{k^2 k^{\prime 2}}{\left(k^2 +\Tilde{m}^2 \right)\left(k^{\prime 2} +\Tilde{m}^2 \right)}=\frac{iM^{\varepsilon}\tilde{m}^2}{\pi}\left[\frac{1}{\varepsilon}+\frac{1}{2}\left(\gamma_E +1\right)+\frac{1}{2}\ln\left(\frac{\Tilde{m}^2}{4\pi M^2}\right)+\mathcal{O}(\varepsilon)\right].  
\end{equation}
Let us now consider the second piece coming from Eq. (\ref{splitting fourth integral}). By writing $k^\prime =k^\prime +\Tilde{m}^2 -\Tilde{m}^2$, we can write this contribution as
\begin{equation}
\int \frac{\mathrm{d}^d k}{(2\pi)^d}\frac{k^{\prime 2}(\Tilde{p}\cdot k)}{\left(k^2 +\Tilde{m}^2 \right)\left(k^{\prime 2}+\Tilde{m}^2 \right)}=\int \frac{\mathrm{d}^d k}{(2\pi)^d}\frac{\Tilde{p}\cdot k}{k^2 +\Tilde{m}^2}- \int \frac{\mathrm{d}^d k}{(2\pi)^d}\frac{\Tilde{m}^2\Tilde{p}\cdot k}{\left(k^2 +\Tilde{m}^2 \right)\left(k^{\prime 2}+\Tilde{m}^2 \right)}.   
\end{equation}
The first integral vanishes since the integrand is antisymmetric under $k\to -k$. Concerning the second one, upon shifting $k\to k+x\Tilde{p}$ and combining the denominator by using Feynman's trick once again, we have
\begin{align}
\int \frac{\mathrm{d}^d k}{(2\pi)^d}\frac{\Tilde{p}\cdot k}{\left(k^2 +\Tilde{m}^2 \right)\left(k^{\prime 2}+\Tilde{m}^2 \right)}&=\int_0^1 dx\int \frac{\mathrm{d}^d k}{(2\pi)^d}\frac{\Tilde{p}\cdot k}{\left[(k-xq)^2 +\Tilde{p}^2 x(1-x)+\Tilde{m}^2 \right]^2}\\
&=\int_0^1 dx\int \frac{\mathrm{d}^d k}{(2\pi)^d} \frac{\Tilde{p}\cdot (k+x\Tilde{p})}{\left[k^2 +\Tilde{p}^2 x(1-x)+\Tilde{m}^2 \right]^2}.
\end{align}
The first term, the one proportional to $\Tilde{p}\cdot k$, vanishes. The remaining one can be written as
\begin{equation}\label{second piece vanishing}
\int \frac{\mathrm{d}^d k}{(2\pi)^d}\frac{\Tilde{p}\cdot k}{\left(k^2 +\Tilde{m}^2 \right)\left(k^{\prime 2}+\Tilde{m}^2 \right)}=\int_0^1 \mathrm{d}x \int \frac{\mathrm{d}^d k}{(2\pi)^d}\frac{x\Tilde{p}^2}{\left[k^2 +\Tilde{p}^2 x(1-x)+\Tilde{m}^2 \right]^2}. 
\end{equation}
We can now proceed in the same way as before, see below Eq. (\ref{before expanding}). However, we immediately notice that the first term of the expansion (\ref{expansion in p}) would be multiplied by $\Tilde{p}^2$, and so we can safely conclude that, in this specific limit, the above integral vanishes. The next contribution in (\ref{splitting fourth integral}) can be also shown to be vanishing. We have:
\begin{align}
\int \frac{\mathrm{d}^d k}{(2\pi)^d}\frac{k^2 (\Tilde{p}\cdot k^\prime)}{\left(k^2 +\Tilde{m}^2 \right)\left(k^{\prime 2}+\Tilde{m}^2 \right)}&=\int \frac{\mathrm{d}^d k}{(2\pi)^d}\frac{\Tilde{p}\cdot k^\prime}{k^{\prime 2}+\Tilde{m}^2}-\int \frac{\mathrm{d}^d k}{(2\pi)^d}\frac{\Tilde{m}^2 \Tilde{p}\cdot k^\prime}{\left(k^2 +\Tilde{m}^2 \right)\left(k^{\prime 2}+\Tilde{m}^2 \right)}\nonumber\\
&=\int \frac{\mathrm{d}^d k}{(2\pi)^d}\frac{\Tilde{p}\cdot k}{k^2 +\Tilde{m}^2}-\int \frac{\mathrm{d}^d k}{(2\pi)^d}\frac{\Tilde{m}^2\Tilde{p}\cdot (k-\Tilde{p})}{\left(k^2 +\Tilde{m}^2 \right)\left(k^{\prime 2}+\Tilde{m}^2 \right)}.
\end{align}
The first term is zero because the integrand is antisymmetric under $k\to -k$. In the second term we recognise two expressions we already proved to be zero in the limit of interest. We now finally consider the last contribution in Eq. (\ref{splitting fourth integral}), which can be written as follows:
\begin{align}
\int \frac{\mathrm{d}^d k}{(2\pi)^d}\frac{(\Tilde{p}\cdot k)(\Tilde{p}\cdot k^\prime)}{\left(k^2 +\Tilde{m}^2 \right)\left(k^{\prime 2}+\Tilde{m}^2 \right)}&=\Tilde{p}_a \Tilde{p}_b \int \frac{\mathrm{d}^d k}{(2\pi)^d}\frac{ k^a (k-\Tilde{p})^b}{\left(k^2 +\Tilde{m}^2 \right)\left(k^{\prime 2}+\Tilde{m}^2 \right)}.
\end{align}
The second term in the numerator of the above expression vanishes since, excluding the factor of $\Tilde{p}^2$, it is exactly the same integral as in Eq. (\ref{second piece vanishing}). Concerning the first piece, we have
\begin{align}
\Tilde{p}_a \Tilde{p}_b \int \frac{\mathrm{d}^d k}{(2\pi)^d}\frac{k^a k^b}{\left(k^2 +\Tilde{m}^2 \right)\left(k^{\prime 2}+\Tilde{m}^2 \right)}&=\Tilde{p}_a \Tilde{p}_b \int_0^1 dx \int \frac{\mathrm{d}^d k}{(2\pi)^d}\frac{(k+x\Tilde{p})^a (k+x\Tilde{p})^b}{\left(k^2 +\Delta \right)^2},
\end{align}
where the definition of $\Delta$ is the same as the one below Eq. (\ref{delta definition}). Splitting the numerator we immediately notice that the above expression gives rise to integrals that vanish as long as $\Tilde{p}\to 0$. Thus, the only non-vanishing contribution in Eq. (\ref{splitting fourth integral}) is the first one. Putting it all together\footnote{Essentially we are considering the prefactors in (\ref{I_4}) as well as the fact that the effective coupling constant has to be written as $M^{-\varepsilon/2}\mu q$.}, we now write down the final result for $I_4$:
\begin{equation}
I_4^d \bigr\rvert_{\Tilde{p}\to 0}=\frac{iM^{-\varepsilon}q^4}{4\pi^3 R^2}\frac{1}{\lambda-1 }\left[\frac{1}{\varepsilon}+\frac{1}{2}\left(\gamma_E +1 \right)+\frac{1}{2}\ln\left(\frac{\lambda-1}{4\pi R^2 M^2}\right)+\mathcal{O}(\varepsilon)\right].
\end{equation}
Thus, summing over all the contributions, $i\mathcal{M}_{\text{seagull}}^{\text{even}}$ in the limit $\varepsilon \to 0$ results in
\begin{equation}\label{final result one-loop four vertex}
i\mathcal{M}_{\text{seagull}}^{\text{even}} = \frac{iq^4}{8\pi^3 R^2}\frac{1}{\lambda -1}.
\end{equation}
The second one-loop diagram with four-vertices arises from the odd mode of the photon and is drawn in \figref{fig:diagram four-vertex 2}. This diagram evaluates to
\begin{align}\label{amplitude odd case}
i\mathcal{M}_{\text{seagull}}^{\text{odd}}&=\int \frac{\mathrm{d}^2 k}{(2\pi)^2}\frac{1}{\lambda -1}\frac{-i}{k^2 +\Tilde{m}^2}\left(-\frac{2iq^2}{4\pi R^2}\right)\frac{1}{\lambda-1}\frac{-i}{k^{\prime 2}+\Tilde{m}^2}\left(-\frac{2iq^2}{4\pi R^2}\right)\nonumber\\
&=\frac{q^4}{4\pi^2 R^4}\frac{1}{\left(\lambda-1\right)^2}\int \frac{\mathrm{d}^2 k}{(2\pi)^2}\frac{1}{k^2 +\Tilde{m}^2}\frac{1}{k^{\prime 2}+\Tilde{m}^2}\\
&=\frac{iq^4}{16\pi^3 R^2}\frac{1}{\left(\lambda-1\right)^3} \, ,
\end{align}
\vspace{0.4cm}
where we made use of the result obtained in (\ref{finite contribution}), in the limit $\varepsilon\to 0$. 

As expected, we see from these results that the four vertex contributions do not scale with the centre of mass energy of the scattering process. They may, nevertheless, be seen as corrections to the eikonal ampltidues (that yield the classical electromagnetic shockwave) that are calculable in this second quantised formalism.

\section{Conclusions}\label{sec:discussion}
In this article, we established an equivalence between the $1 \rightarrow 1$ S-matrix in the first quantised formalism arising from electromagnetic shockwaves as classical solutions to Maxwell equations near the black hole horizon and the $t$-channel elastic $2\rightarrow 2$ in the black hole eikonal limit. In order to do so, we developed a second quantised theory for electromagnetic fluctuations and charged particle scattering near the black hole horizon. While the $1\rightarrow 1$ result builds on \cite{tHooft:1996rdg}, the $2\rightarrow 2$ result extends the formalism first developed in \cite{Gaddam:2020rxb, Gaddam:2020mwe}.

The formalism developed in this article is naturally suited for incorporating other forces of the Standard Model. It would be very interesting to see if there are non-Abelian shockwaves near the horizon and if new physics emerges. The second quantised theory allows for a calculation of various quantities of physical interest, including corrections to the electromagnetic potential near the horizon in the spirit of \cite{Donoghue:1993eb} and other classical observables \cite{Bern:2021xze}. The gravitational eikonal also led to speculation about a certain antipodal correlation on the bifurcation sphere on the horizon \cite{Hooft:2016itl, Betzios:2017krj, Betzios:2020wcv}. It would be interesting to find an electromagnetic analog of these proposals. An analog of the relation between the shockwave algebra and the soft algebra near null-infinity found in \cite{He:2023qha} is also an interesting question to explore near the horizon of a black hole. 

In \cite{Aggarwal:2023qwl}, all symmetries associated with the near-horizon scattering of gravitational radiation have been derived. The techniques developed there can easily be adapted to the electromagnetic radiation that will emerge from the theory developed here in this paper. Such electromagnetic radiation is expected to result in a near-horizon memory effect that may have observable consequences in the spectral fluctuations of stellar oscillations as discussed in \cite{Aggarwal:2023qwl}.

\section*{Acknowledgements}
We are grateful to Gerard 't Hooft for various helpful conversations over the years. We acknowledge the support of the Netherlands Organisation for Scientific Research (NWO) and the Delta-Institute for Theoretical Physics (D-ITP) that is funded by the Dutch Ministry of Education, Culture and Science (OCW). Nava G. is currently supported by project RTI4001 of the Department of Atomic Energy, Govt. of India.

\appendix

\section{Maxwell theory in partial waves}\label{app:MaxwellActionPW}
The quadratic Lagrangian for the Maxwell field written in terms of \eqref{eqn:quadraticOperPhoton1} is
\begin{align}\label{eqn:quadraticOperPhoton2}
\mathcal{L}_\gamma =\frac{1}{2}A_\mu \mathcal{O}^{\mu\nu}A_\nu =\frac{1}{2}\left(A_a \mathcal{O}^{ab}A_b +A_a \mathcal{O}^{aB}A_B +A_B \mathcal{O}^{Ba}A_a +A_A \mathcal{O}^{AB}A_B\right),  
\end{align}
where the four terms above are given by
\begin{align}
A_a \mathcal{O}^{aB}A_B &= -A_a^+ \Tilde{\nabla}^a \left(V^b A_b^+ \right), \label{L1}\\
A_B \mathcal{O}^{Ba}A_a &= \frac{1}{2}g^{AB}A_A^- \left(V_a \Tilde{\nabla}^a -V_a V^a \right)A_B^- , \label{L2}\\
A_A \mathcal{O}^{AB}A_B &= A_A^- \left(g^{AB}\hat{\nabla}^C \hat{\nabla}_C -\frac{1}{4}g^{AB}V^a V_a +g^{AB}\tilde{\square} \right)A_B^- -\frac{1}{2}A_A^- g^{AB}\Tilde{\nabla}^a \left(V_a A_B^- \right), \label{L3}\\
A_a \mathcal{O}^{ab}A_b &= A_a^+ \left(g^{ab}\Tilde{\square} -g^{ac}g^{bd}\tilde{\nabla}_c \tilde{\nabla}_d +\frac{1}{r^2}g^{ab}\Delta_\Omega +g^{ab}V^d \tilde{\nabla}_d -\frac{1}{2}g^{ac}V_c V^b \right)A_b^+. \label{L4}
\end{align}
Above, the differential operators $\Tilde{\nabla}$ and $\Tilde{\square}$ are the covariant derivative and the d'Alembertian on the light-cone, respectively. Lower-case Latin indices are raised and lowered with the light-cone metric $g_{ab}$. Moreover, $\hat{\nabla}$ represents the covariant derivative on the sphere, and upper-case Latin indices are raised and lowered with the metric of the two-sphere $g_{AB}$. Finally, we denoted the Laplacian on the unit round sphere with $\Delta_\Omega$ and defined the vector potential $V_a \coloneqq 2\partial_a \log r$. From the above expressions, we immediately notice the decoupling between even- and odd-parity modes. These expressions are derived by direct computation, following the strategy laid out in the gravitational case in \cite{Gaddam:2020rxb, Gaddam:2020mwe}.

We would like to employ the partial wave decomposition described in \secref{sec:QEDhorizon}. To this end, we first notice that the sum of \eqref{L1} and \eqref{L4} gives
\begin{equation}
A_a \mathcal{O}^{ab}A_b +A_a \mathcal{O}^{aB}A_B =\sum_{\ell,m;\ell^\prime ,m^\prime}Y_{\ell m}Y_{\ell^\prime m^\prime}A_{\ell m,a}\mathcal{P}^{ab}A_{\ell^\prime m^\prime ,b},    
\end{equation}
where the operator $\mathcal{P}^{ab}$ is given by
\begin{equation}
\mathcal{P}^{ab} \coloneqq g^{ab}\tilde{\square}-\Tilde{\nabla}^a \Tilde{\nabla}^b -g^{ab}\frac{\lambda -1}{r^2}+g^{ab}V^c \tilde{\nabla}_c -\frac{1}{2}V^a V^b -\Tilde{\nabla}^a V^b -V^b \Tilde{\nabla}^a.   
\end{equation}
To arrive at this expression, we have used that $\Delta_\Omega Y_{\ell m} = \ell(\ell+1)Y_{\ell m} \coloneqq \left(\lambda - 1 \right) Y_{\ell m}$. Next, the sum of \eqref{L2} and \eqref{L3} gives
\begin{equation}
A_B \mathcal{O}^{Ba}A_a +A_A \mathcal{O}^{AB}A_B = \sum_{\ell,m;\ell^\prime ,m^\prime}g^{AB}\left(\epsilon_A^{\ \ C} \partial_C Y_{\ell m}\right) \left(\epsilon_B^{\ \ D} \partial_D Y_{\ell^\prime m^\prime}\right) A_{\ell m,2}\mathcal{P}A_{\ell^\prime m^\prime,2},
\end{equation}
where the operator $\mathcal{P}$ is now
\begin{equation}
\mathcal{P} \coloneqq \tilde{\square}+\frac{2-\lambda}{r^2}-\frac{1}{2}\Tilde{\nabla}^a V_a -\frac{3}{4}V_a V^a.
\end{equation}
The resulting action for the even mode of the photon is therefore
\begin{align}
S_{\gamma,\text{even}} & \coloneqq \frac{1}{2}\sum_{\ell,m}\int \mathrm{d}^2 x A(r)r^2 A_{\ell m,a}\mathcal{P}^{ab}A_{\ell^\prime m^\prime ,b},
\end{align}
where we used the usual orthogonality of the scalar spherical harmonics. Similarly, the action for the odd-parity contribution is given by
\begin{align}
S_{\gamma,\text{odd}}&=\frac{1}{2}\sum_{\ell,m;\ell^\prime ,m^\prime}\int \mathrm{d}\Omega g^{AB}\left(\epsilon_A^{\ \ C} \partial_C Y_{\ell m}\right) \left(\epsilon_B^{\ \ D} \partial_D Y_{\ell^\prime m^\prime}\right)\int \mathrm{d}^2 x A(r)r^2 A_{\ell m,2}\mathcal{P}A_{\ell^\prime m^\prime ,2}\nonumber\\
&=\frac{1}{2}\sum_{\ell,m}\int \mathrm{d}^2 x A(r)r^2 A_{\ell m,2}\mathcal{P}A_{\ell^\prime m^\prime ,2} \, .
\end{align}
where this time we used the orthogonality relation for the vector spherical harmonics
\begin{equation}
\int \mathrm{d}\Omega g^{AB}\left(\epsilon_A^{\ \ C} \partial_C Y_{\ell m}\right) \left(\epsilon_B^{\ \ D} \partial_D Y_{\ell^\prime m^\prime}\right)  =\left(\lambda -1 \right)\delta_{\ell \ell^\prime}\delta_{mm^\prime} \, ,
\end{equation}
and absorbed the factor $\lambda - 1$ into the operator. Spherical symmetry of the background has ensured that the four-dimensional theory has been reduced to an infinite tower of decoupled two-dimensional theories, one for each partial wave. In order to write down the action in a form suitable for our purposes, we will absorb a factor of $r^2$ into the fields; this is achieved by introducing the following new operator acting on the fields:
\begin{equation}
\mathcal{D}_a (\boldsymbol{\cdot}) \coloneqq \Tilde{\nabla}_a (\boldsymbol{\cdot}) +\frac{1}{2}V_a (\boldsymbol{\cdot}) =\frac{1}{r}\Tilde{\nabla}_a (r\hspace{0.1cm}\boldsymbol{\cdot}) \, .
\end{equation}
The operators $\mathcal{P}_{ab}$ and $\mathcal{P}$ can then be written in terms of $\mathcal{D}_a$ as
\begin{align}
\mathcal{P}^{ab} &= g^{ab}\left[\mathcal{D}_a \mathcal{D}^a -F_c^c -\frac{\lambda -1}{r^2}\right]-\mathcal{D}^a \mathcal{D}^b -F^{ab}-V^{[b}\mathcal{D}^{a]}, \\  
\mathcal{P} &= \left(\lambda -1\right) \mathcal{D}_a \mathcal{D}^a +\frac{\left(\lambda -1\right) \left(2-\lambda \right)}{r^2}-2\left(\lambda -1\right) F_a^a -\left(\lambda -1\right) V_a \mathcal{D}^a,
\end{align}
where we have defined the symmetric tensor $F_{ab}$ as
\begin{equation}\label{symmetric tensor F}
F_{a b}:=\frac{1}{2} \mathcal{D}_{(a} V_{b)}=\frac{1}{r} \tilde{\nabla}_{a} \tilde{\nabla}_{b} r =\frac{1}{2} \tilde{\nabla}_{(a} V_{b)}+\frac{1}{4} V_{a} V_{b}.
\end{equation}
From the definition of $\mathcal{D}_a$, it is straightforward to derive the identities
\begin{align}
\mathcal{D}^a \mathcal{D}_a (\boldsymbol{\cdot}) &=\frac{1}{r}\Tilde{\square}(r\hspace{0.1cm}\boldsymbol{\cdot}), \label{relation 1} \\
\mathcal{D}^a \mathcal{D}^b (\boldsymbol{\cdot})&=\frac{1}{r}\Tilde{\nabla}^a \Tilde{\nabla}^b (r\hspace{0.1cm}\boldsymbol{\cdot}), \label{relation 2} \\
V^{[b}\mathcal{D}^{a]}(\boldsymbol{\cdot})&=\frac{1}{r}V^{[b}\mathcal{\Tilde{\nabla}}^{a]}(r\hspace{0.1cm}\boldsymbol{\cdot}), \label{relation 3}
\end{align}
which allow us to rewrite the two integrands in the even and odd actions (ignoring the function $A(r)$ for the moment) as\footnote{For simplicity, we suppressed the partial wave indices.}
\begin{align}
r^2 A_a \mathcal{P}^{ab}A_b &\xrightarrow{(\ref{relation 1})-(\ref{relation 3})}rA_a \left[g^{ab}\left(\Tilde{\square}-F_c^c -\frac{\lambda -1}{r^2}\right)-\Tilde{\nabla}^a \Tilde{\nabla}^b -F^{ab}-V^{[b}\Tilde{\nabla}^{a]}\right]rA_b, \\
r^2 A_2 \mathcal{P}A_2 &\xrightarrow{(\ref{relation 1})-(\ref{relation 3})}rA_2 \left[\left(\lambda -1\right) \left(\Tilde{\square}-2F_a^a -V_a \Tilde{\nabla}^a\right)+\frac{\left(\lambda -1\right) \left(2-\lambda \right)}{r^2} \right]rA_2.
\end{align}
We may now safely make the following field redefinitions:
\begin{equation}
\Tilde{A}_a \coloneqq rA_a, \quad \mathcal{A}\coloneqq rA_2.    
\end{equation}
Th complete quadratic action for the photon can then be written as
\begin{equation}
S_\gamma = S_{\gamma,\text{even}}+S_{\gamma,\text{odd}}=\frac{1}{2}\int \mathrm{d}^2 x\sqrt{-\Tilde{g}}\Tilde{A}^a \Tilde{\Delta}_{ab}^{-1}\Tilde{A}^b +\frac{1}{2}\int \mathrm{d}^2 x\sqrt{-\Tilde{g}}\Tilde{A}\Tilde{\Delta}^{-1}\Tilde{A},
\end{equation}
where $\sqrt{-\det \left(g_{ab}\right)} \coloneqq \sqrt{-\tilde{g}} = A(r)$, while the operators $\Tilde{\Delta}_{ab}^{-1}$ and $\Tilde{\Delta}^{-1}$ are given by
\begin{align}
\Tilde{\Delta}_{ab}^{-1}&\coloneqq g_{ab}\left(\Tilde{\square}-F_c^c -\frac{\lambda -1}{r^2}\right)-\Tilde{\nabla}_a \Tilde{\nabla}_b -F_{ab}-V_{[b}\Tilde{\nabla}_{a]}, \label{even-parity operator}\\
\Tilde{\Delta}^{-1}&\coloneqq \left(\lambda -1\right) \Tilde{\square}+\frac{\left(\lambda -1\right)\left(2-\lambda \right)}{r^2}-2\left(\lambda -1\right) F_a^a -\left(\lambda -1\right) V_a \Tilde{\nabla}^a. \label{odd-parity operator}
\end{align}
To find the photon propagator, we observe that the metric in the effective two-dimensional theory is conformally flat, i.e, $\Tilde{g}_{ab}=A(r)\eta_{ab}$, where $\eta_{ab}$ is the two-dimensional Minkowski metric in light-cone coordinates with off-diagonal elements being $-1$. This allows us to rewrite the theory in flat space with curvature effects traded for potentials. As an illustration, consider the action for the odd mode:
\begin{align}
S_{\gamma,\text{odd}} &= \frac{\lambda -1}{2}\int \mathrm{d}^2 xA(r)\mathcal{A}\left[\tilde{g}^{ab}\Tilde{\nabla}_a \tilde{\nabla}_b +\frac{2-\lambda}{r^2}-2\tilde{g}^{ab}F_{ab}-\Tilde{g}^{ab}V_a \tilde{\nabla}_b \right]\mathcal{A} \nonumber\\
&= \frac{\lambda -1}{2}\int \mathrm{d}^2 xA(r)\mathcal{A}\left[\frac{\eta^{ab}}{A(r)}\Tilde{\nabla}_a \tilde{\nabla}_b +\frac{2-\lambda}{r^2}-\frac{2\eta^{ab}}{A(r)}F_{ab}-\frac{\eta^{ab}}{A(r)}V_a \tilde{\nabla}_b\right]\mathcal{A},
\end{align}
Pulling the function $A\left(r\right)$ through the quadratic operator, the odd action becomes:
\begin{equation}
S_{\gamma,\text{odd}} = \frac{\lambda -1}{2}\int \mathrm{d}^2 x \mathcal{A}\left[\partial^2 +A(r)\frac{2-\lambda}{r^2}-2F_a^a -V^b \partial_b \right]\mathcal{A},   
\end{equation}
where we defined $\partial^2 \coloneqq\eta^{ab}\partial_a \partial_b $. Furthermore, we need to consistently redefine the antisymmetric tensor $F_{ab}$ in terms of partial derivatives. From it definition \eqref{symmetric tensor F}, we have
\begin{equation}\label{F_{ab}}
F_{ab}=\frac{1}{4}\Tilde{\nabla}_a V_b +\frac{1}{4}\Tilde{\nabla}_b V_a +\frac{1}{4}V_a V_b =\frac{1}{4}\left(\partial_a V_b -\Gamma_{ab}^e V_e \right)+\frac{1}{4}\left(\partial_b V_a -\Gamma_{ba}^e V_e \right)+\frac{1}{4}V_a V_b .
\end{equation}
We now express the Christoffel symbols of the form $\Gamma_{ab}^e$ as
\begin{equation}
\Gamma_{ab}^e =2\delta_{(a}^e U_{b)}-\tilde{g}_{ab}U^e =2\delta_{(a}^e U_{b)}-\tilde{g}_{ab}\tilde{g}^{de}U_d =2\delta_{(a}^e U_{b)}-\eta_{ab}\eta^{de}U_d =2\delta_{(a}^e U_{b)}-\eta_{ab}U^e,
\end{equation}
where we introduced a new potential, $U_a$, defined as
\begin{equation}
U_a \coloneqq\frac{1}{2A(r)}\partial_a A(r).
\end{equation}
Therefore, we have that
\begin{align}
F_{ab} &= \frac{1}{2}\partial_{(a}V_{b)}-\frac{1}{2}\left(\delta_a^e U_b +\delta_b^e U_a -\eta_{ab}U^e \right)V_e +\frac{1}{4}V_a V_b \nonumber\\
&= \frac{1}{2}\partial_{(a}V_{b)}-\frac{1}{2}U_b V_a -\frac{1}{2}U_a V_b +\frac{1}{2}\eta_{ab}U^e V_e +\frac{1}{4}V_a V_b \nonumber\\
&= \frac{1}{2}\partial_{(a}V_{b)}-U_{(a}V_{b)}+\frac{1}{2}\eta_{ab}U^e V_e +\frac{1}{4}V_a V_b.
\end{align}
The last equality in the above expression is our new definition of $F_{ab}$, after the rescaling. An analogous procedure yields the following action for the even photon:
\begin{equation}
S_{\gamma,\text{even}} = \frac{1}{2}\int \mathrm{d}^2 xA(r)\Tilde{A}^a \left[g_{ab}\left(\Tilde{\square}-F_c^c -\frac{\lambda -1}{r^2}\right)-\Tilde{\nabla}_a \Tilde{\nabla}_b -F_{ab}-V_{[b}\Tilde{\nabla}_{a]}\right]\Tilde{A}^b.
\end{equation}
Redefining $\tilde{A}_a \coloneqq\sqrt{A(r)}\mathcal{A}_a$ and using that $\tilde{g}_{ab} = A(r)\eta_{ab}$, we obtain
\begin{multline}
S_{\gamma,\text{even}}=\frac{1}{2}\int \mathrm{d}^2 x\sqrt{A(r)}\mathcal{A}^a \left[\eta_{ab}\left(\eta^{cd}\Tilde{\nabla}_c \Tilde{\nabla}_d -\eta^{cd}F_{cd}-A(r)\frac{\lambda -1}{r^2}\right) \right.\\
\left. -\Tilde{\nabla}_a \Tilde{\nabla}_b -F_{ab}-\frac{1}{2}V_b \Tilde{\nabla}_a +\frac{1}{2}V_a \Tilde{\nabla}_b \right]\frac{1}{\sqrt{A(r)}}\mathcal{A}^b .
\end{multline}
Evaluating the action of the operator in square brackets on $\mathcal{A}^b /\sqrt{A(r)}$, we find
\begin{multline}
S_{\gamma,\text{even}} = \frac{1}{2}\int \mathrm{d}^2 x\mathcal{A}^a \left[\eta_{ab}\left(\partial^2 -U_c U^c +\frac{1}{2}V_c U^c -\frac{1}{2}\partial_c V^c -\frac{1}{4}V_c V^c -A(r)\frac{\lambda -1}{r^2}\right) \right.\\
\left. +\hspace{0.06cm}2U_{[b}\partial_{a]}+U_a U_b -\partial_a U_b -\partial_a \partial_b -V_{[b}\partial_{a]}-F_{ab}\right]\mathcal{A}^b.
\end{multline}
To sum up, the complete quadratic photon action $S_\gamma$ is now given by
\begin{equation}\label{final action photon}
S_\gamma = S_{\gamma,\text{even}}+S_{\gamma,\text{odd}} = \frac{1}{2}\int \mathrm{d}^2 x\mathcal{A}^a \Delta_{ab}^{-1}\mathcal{A}^b +\frac{1}{2}\int \mathrm{d}^2 x\mathcal{A}\Delta^{-1}\mathcal{A},
\end{equation}
where the operators after the rescaling have been defined as
\begin{align}
\begin{split}\label{final operator 1}
\Delta_{ab}^{-1} \coloneqq{} &\eta_{ab}\left(\partial^2 -U_c U^c +\frac{1}{2}V_c U^c -\frac{1}{2}\partial_c V^c -\frac{1}{4}V_c V^c -A(r)\frac{\lambda -1}{r^2}\right) \\
&\hspace{2cm}+2U_{[b}\partial_{a]}+U_a U_b -\partial_a U_b -\partial_a \partial_b -V_{[b}\partial_{a]}-F_{ab},
\end{split} \\[0.8ex]
\begin{split}\label{final operator 2}
\Delta^{-1} \coloneqq{} & \left(\lambda -1 \right) \partial^2 +A(r)\frac{\left(\lambda -1 \right) \left(2-\lambda\right)}{r^2}-2\left(\lambda -1 \right) F_a^a -\left(\lambda -1 \right) V^b \partial_b.
\end{split}
\end{align}
From their definitions, the potentials appearing in these quadratic operators satisfy:
\begin{align}
V_a &= \frac{A}{rR}x_a ,\label{V_a}\\
U_a &= -\frac{A}{4rR}\left(1+\frac{r}{R}\right)x_a,\\
\partial_a V_b &=\frac{A}{rR}\eta_{ab}-\frac{A^2}{2R^2 r^2}\left(2+\frac{r}{R}\right)x_a x_b,\\
\partial_a U_b &= -\frac{A}{4rR}\left(1+\frac{r}{R}\right) \eta_{ab}+\frac{A^2}{8R^2 r^2}\left(2+2\frac{r}{R}+\frac{r^2}{R^2}\right)x_a x_b,\\
F_{ab} &= \frac{AR}{2r^3}\eta_{ab},\label{F_ab}
\end{align}
where we recall that the Schwarzschild background is specified by
\begin{equation}\label{A and UV Schw background}
A(r) = \frac{R}{r}e^{1-\frac{r}{R}}, \quad UV = 2R^2 \left(1-\frac{r}{R}\right)e^{\frac{r}{R}-1}.
\end{equation}
By inserting the above expressions in Eqs. (\ref{final operator 1}) and (\ref{final operator 2}), we obtain
\begin{subequations}
\begin{align}
\Delta_{ab}^{-1} &= \eta_{ab} \left\{ \eta^{c d} \partial_{c} \partial_{d} -\dfrac{A\left(r\right)^2}{16 r^2 R^2} \left[ \left( 1+ \dfrac{r}{R} \right)^2 + 2 \left( 1 + \dfrac{r}{R}\right) - 8 \left( 2 + \dfrac{r}{R} \right) + 4 \right] x_a x^a - \dfrac{A\left(r\right)}{rR} \right. \nonumber \\
& \qquad \qquad \left. -\hspace{1mm} \dfrac{\left(\lambda - 1\right) A\left(r\right)}{r^2} + \dfrac{A\left(r\right)}{4 r R} \left( 1 + \dfrac{r}{R} \right) - \dfrac{A\left(r\right) R}{2 r^3}\right\} - \dfrac{A\left(r\right)}{4 r R} \left( 3 + \dfrac{r}{R} \right) \left(x_b \partial_a - x_a \partial_b \right) \nonumber \\
& \qquad \qquad + \dfrac{A\left(r\right)^2}{16 r^2 R^2} \left[ \left( 1 + \dfrac{r}{R} \right)^2 - 2 \left( 2 + 2 \dfrac{r}{R} + \dfrac{r^2}{R^2} \right) \right] x_a x_b  - \partial_{a} \partial_{b}\label{eqn:generalEvenOper} \, , \\
\Delta^{-1} &= \left(\lambda - 1\right) \eta^{a b} \partial_{a} \partial_{b} - \dfrac{A\left(r\right) \left(\lambda - 1\right) \left( \lambda - 2\right)}{r^{2}} - \dfrac{2 \left(\lambda - 1\right) A\left(r\right) R}{r^{3}} \nonumber \\
	&\qquad \qquad  - \dfrac{A\left(r\right) \left(\lambda - 1\right) }{ r R} x^{a} \partial_{a} \, , \label{eqn:generalOddOper}
\end{align}
\end{subequations}
which is the result quoted in the main text in \eqref{eqn:generalQuadOpers}.

\section{A certain factor of $4\pi$}\label{app:4pi}
As can be seen from comparing \eqref{final expression general loop amplitude pp case} in the field-theoretical eikonal and \eqref{eqn:shockwaveQMSmatrix} of the first quantised shockwave calculation, there is a curious discrepancy of a factor of $4\pi$. This factor arises in the field theory calculation from the $s$-wave component of one of the scalars in the three-point vertex in \secref{sec:threeVertex}. One may suspect that the difference between the two resides in the sources. Should the charge densities in the two frameworks be the same, we expect the S-matrix elements to agree. In what follows, we show that this is indeed true.

In the quantum-mechanics case, the $v$-component of the current density, in a given partial, can be directly read off from Eq. \eqref{eqn:chargeDensityQM}:
\begin{equation}
J_v^{\ell m} ~ = ~ - \dfrac{q_{\text{in}}^{\ell m}}{R^{2}} \delta(u) \, .    
\end{equation}
On the other hand, in the field-theory side, the $v$-component of the current is given by
\begin{equation}\label{current density QM}
j_v ~ = ~ - i q_{\text{in}} \left(\phi^* \partial_v \phi -v\phi \partial_v \phi^* \right) \, ,
\end{equation}
as can be seen from \eqref{eqn:current}. Expanding in spherical harmonics, inserting the rescaling $\phi= \varphi/R$, and demanding one of the scalars to be in the $s$-wave, we find
\begin{equation}\label{current density QFT}
j_v^{\ell m} ~ = ~ - i \dfrac{q_{\text{in}}}{R^2} Y_{00} \left( \varphi_{0}^* \partial_v \varphi_{\ell m} - \varphi_{\ell m} \partial_v \varphi_{0}^* + \varphi_{\ell m}^* \partial_v \varphi_{0} - \varphi_{0} \partial_v \phi_{\ell m}^* \right) \, .
\end{equation}
The main difference between \eqref{current density QM} and \eqref{current density QFT} is that the current density in quantum field theory is an operator. Thus, a proper comparison warrants its expectation value of $j_v^{\ell m}$ in an appropriately defined initial state:
\begin{equation}\label{initial state}
|\text{in} \rangle ~ = ~ \int \dfrac{\mathrm{d} p}{2 \pi} \Phi\left(p\right) \times \dfrac{1}{\sqrt{2}} \left(a^{\dagger}_{\ell m}\left(p\right) + a^{\dagger}_{0}\left(p\right)\right) |0\rangle \, ,
\end{equation}
where $\Phi(p)$ is a normalized test function localized around a specific momentum, say, $p=p_1$. The expression above, \eqref{initial state}, represents a one-particle state where we have a superposition of two shells at equal momentum where one of them is in the $s$-wave. We now recall that the only non-vanishing commutation relations are 
\begin{equation}
\left[a_{\ell m} \left(p\right) , a^{\dagger}_{\ell^\prime m^\prime} \left(p^\prime \right)\right] ~ = ~ 2 \pi \delta\left(p - p^\prime \right) \delta_{\ell \ell^\prime}\delta_{m m^\prime} \, .
\end{equation}
To compute the expectation value of $j_{v}^{\ell m}$, we begin with the first term in \eqref{current density QM}:
\begin{align}
- i \dfrac{q_{\text{in}}}{R^2} Y_{00} \langle \text{in} | \varphi_{0}^* \partial_v \varphi_{\ell m} | \text{in} \rangle ~ &= ~ - \dfrac{1}{2 R^{2}} Y_{00} \int \dfrac{\mathrm{d} p \mathrm{d} p^\prime}{\left(2 \pi\right)^2} \dfrac{p^\prime}{\sqrt{p p^\prime}} \int \dfrac{\mathrm{d} k \mathrm{d} k^\prime}{ \left(2 \pi\right)^2} \Phi^* \left(k\right)\Phi\left(k^\prime\right) \nonumber \\
&\qquad \times \langle 0 | a_{0}\left(k\right) a_{\ell m} \left(p\right) a_{\ell m}^\dagger \left(p^\prime \right) a_{\ell m}^\dagger \left(k^\prime \right) | 0 \rangle e^{ i (p - p^\prime) x} \nonumber \\
&= ~ - \dfrac{1}{4 R^{2}}Y_{00} \int \dfrac{\mathrm{d} p \mathrm{d} p^\prime}{\left(2 \pi\right)^2} \dfrac{p^\prime}{\sqrt{p p^\prime}} \int \dfrac{\mathrm{d}k \mathrm{d} k^\prime}{\left(2 \pi\right)^2} \Phi^* \left(k\right) \Phi \left(k^\prime\right) \nonumber \\
&\qquad \times \left(2 \pi\right)^2 \delta\left(k - p\right) \delta\left(k^\prime - p^\prime \right) e^{ i (p - p^\prime ) x} \, . 
\end{align}
Similar expressions follow for the second and third terms in \eqref{current density QFT} and we may write the complete expectation value as\footnote{For a state $\Phi\left(p\right)$ with support sharply localised in momentum at $p_{1}$, the contribution to the integral from polynomials essentially comes from $p' = p = p_{1}$ allowing us to kill the polynomial pre-factors. The exponentials however, oscillate faster and need to be kept.}
\begin{align}
\langle \text{in} | j_v^{\ell m} | \text{in} \rangle ~ &= ~ - \dfrac{1}{2 R^{2}} q_{\text{in}}Y_{00} \int \dfrac{\mathrm{d} p \mathrm{d} p^\prime}{\left(2 \pi\right)^2} \dfrac{p^\prime}{\sqrt{p p^\prime}} \int \dfrac{\mathrm{d} k \mathrm{d} k^\prime}{\left(2 \pi\right)^2} \Phi^* \left(k\right) \Phi\left(k^\prime\right) \nonumber\\
&\quad\quad \times \left(2 \pi\right)^2 \left[\delta\left( k - p\right) \delta\left(k^\prime - p^\prime \right) e^{ i (p - p^\prime ) x }+ \delta\left(k - p^\prime \right) \delta\left(p - k^\prime \right) e^{ - i ( p - p^\prime ) x} \right] \nonumber \\
&= ~ - \dfrac{q_{\text{in}}}{2 R^{2}} Y_{00} \int \dfrac{\mathrm{d} p \mathrm{d} p^\prime}{\left(2 \pi\right)^2} \left(\Phi^* \left( p \right) \Phi\left( p^\prime \right) e^{ i ( p - p' ) x} + \Phi^* \left( p^\prime \right) \Phi\left(p\right) e^{ - i ( p - p' ) x}\right) \nonumber \\
&= ~ - \dfrac{q_{\text{in}}}{R^{2}}Y_{00} \left|\Phi\left(x\right) \right|^2 \, ,
\end{align}
where we made use of the inverse Fourier transform
\begin{equation}
\Phi\left(x\right) ~ = ~ \int \dfrac{\mathrm{d} p}{2 \pi} \Phi\left(p\right) e^{- i p x} \, . 
\end{equation}
Now, interpreting $|\Phi(x)|^2$ as a probability distribution, we assume the particle with momentum $p_1$ is localized at $u=0$, and find
\begin{equation}\label{expectation value current density}
\langle J_v^{\ell m} \rangle ~ = ~ - \dfrac{q_{\text{in}}}{R^{2}} Y_{00} \delta(u) ~ = ~ - \dfrac{q_{\text{in}}}{\sqrt{4\pi} \, R^{2}} \delta(u) \, .    
\end{equation}
Comparing \eqref{current density QM} and \eqref{expectation value current density}, we see that $q_{in}^{lm} = q_{in}/\sqrt{4\pi}$. This resolves the apparent discrepancy between the field theory and shockwave results, leading to perfect agreement.

\printbibliography

\end{document}